\documentclass[10pt,english,letterpaper]{article}
\usepackage[T1]{fontenc}
\usepackage[latin9]{inputenc}
\usepackage{geometry}
\usepackage{babel}
\usepackage{amsmath}
\usepackage{amssymb}
\usepackage{graphicx}
\usepackage{setspace}
\usepackage{esint}
\usepackage[unicode=true,pdfusetitle,
 bookmarks=true,bookmarksnumbered=true,bookmarksopen=true,bookmarksopenlevel=3,
 breaklinks=false,pdfborder={0 0 0},backref=false,colorlinks=false]
 {hyperref}

\makeatletter

\providecommand{\tabularnewline}{\\}

\usepackage[T1]{fontenc}
\usepackage{ae}
\usepackage{aecompl}

\usepackage[scaled=0.92]{helvet}

\usepackage[sf,bf,small]{titlesec}

\usepackage[small,sf,bf,hang]{caption}

\usepackage[multiple]{footmisc}
\long\def\symbolfootnote[#1]#2{\begingroup%
\def\thefootnote{\fnsymbol{footnote}}\footnote[#1]{#2}\endgroup}

\newcommand{\tablespacer}{\vspace{1mm}}

\usepackage{titletoc}
\dottedcontents{section}[6mm]{\smallskip }{6mm}{0pc} 
\titlecontents*{subsection}[6mm]{\itshape\small }{\thecontentslabel \ }{}{, \thecontentspage}[ \ \textbullet \ \ ][] 
\def\tableofcontents{\subsection*{\contentsname}\vspace{-2mm}\@starttoc{toc}}

\usepackage[nosort]{cite}

\newcommand{\frp}{[RP]}
\newcommand{\fbig}{[\mathrm{Big}]}

\renewcommand{\bar}[1]{\overline{#1}}

\DeclareMathOperator{\re}{Re}  

\newcommand{\res}{\mathop{\mathrm{Res}}} 

\AtBeginDocument{
  
}

\makeatother

\begin{document}
\begin{flushright}
TIFR/TH/11-46\\
arXiv:1111.2839v3\bigskip
\par\end{flushright}

\begin{center}
\textsf{\textbf{\Large Real and Virtual Bound States in \smallskip\smallskip\smallskip
}}\\
\textsf{\textbf{\Large L\"{u}scher Corrections for $CP^{3}$ Magnons}}
\par\end{center}{\Large \par}

\begin{singlespace}
\begin{center}
\vspace{-2mm}Michael C. Abbott$,^{1}$ In\^{e}s Aniceto$^{2}$ and Diego Bombardelli$^{3}$\emph{
}\bigskip \\
{\small $^{1}$}\emph{\small{} Tata Institute of Fundamental Research, }\\
\emph{\small Homi Bhabha Rd, Mumbai 400-005, India}\\
\emph{\small abbott@theory.tifr.res.in }{\small \bigskip }\emph{\small }\\
{\small $^{2}$}\emph{\small{} CAMGSD, Departamento de Matem\'{a}tica, Instituto
Superior T\'{e}cnico, }\\
\emph{\small Av. Rovisco Pais, 1049-001 Lisboa, Portugal}{\small }\\
\emph{\small ianiceto@math.ist.utl.pt}{\small \bigskip }\emph{\small }\\
{\small $^{3}$}\emph{\small{} Centro de F\'{i}sica do Porto and Departamento de
F\'{i}sica e Astronomia,}\\
\emph{\small{} Faculdade de Ciências da Universidade do Porto,}{\small }\\
{\small{} }\emph{\small Rua do Campo Alegre 687, 4169-007 Porto, Portugal}\\
 \emph{\small diego.bombardelli@fc.up.pt}{\small{} }
\par\end{center}{\small \par}
\end{singlespace}

\begin{center}
\bigskip  11 November 2011, with additions 31 May 2012.
\par\end{center}

\subsection*{\hspace{9mm}Abstract}
\begin{quote}
We study classical and quantum finite-size corrections to giant magnons in $AdS_{4}\times CP^{3}$
using generalised L\"{u}scher formulae. L\"{u}scher F-terms are organised in powers
of the exponential suppression factor $(e^{-\Delta/2h})^{m}$, and we calculate all
terms in this series, matching one-loop algebraic curve results from our previous
paper \cite{Abbott:2010yb}. Starting with the second term, the structure of these
terms is different to those in $AdS_{5}\times S^{5}$ thanks to the appearance of
heavy modes in the loop, which can here be interpreted as two-particle bound states
in the mirror theory. By contrast, physical bound states can represent dyonic giant
magnons, and we also calculate F-terms for these solutions. L\"{u}scher $\mu$-terms,
suppressed by $e^{-\Delta/E}$, instead give at leading order the classical finite-size
correction. For the elementary dyonic giant magnon we recover the correction given
by \cite{Abbott:2009um}. We then extend this to calculate the next term in $1/h$,
giving a one-loop prediction. Finally we also calculate F-terms for the various composite
giant magnons, $RP^{3}$ and `big', again finding agreement to all orders. 

\bigskip  \thispagestyle{empty} 
\end{quote}
\tableofcontents{}

\section{Introduction}

The L\"{u}scher formulae arise from one-loop Feynman diagrams in which a propagator
circles the space \cite{Luscher:1983rk,Luscher:1985dn,Klassen:1990ub}. Since such
diagrams are the only ones not present in infinite volume, they give finite-volume
corrections. And because the large-volume limit puts this propagator on-shell, they
can be written in terms of the S-matrix for asymptotic states. 

In the AdS/CFT context, the relevant spatial circle is either the position along
the spin chain (at weak coupling) or else on the worldsheet of a closed string (at
strong coupling), and the S-matrix is known for all values of the coupling \cite{Staudacher:2004tk,Beisert:2005tm,Beisert:2006qh,Arutyunov:2006yd}.
At weak coupling, L\"{u}scher corrections can be successfully summed to reach all
the way to the shortest nontrivial chain, the Konishi operator \cite{Janik:2010kd}.
In this paper however we will be working on the string side of the correspondence,
and following \cite{Ambjorn:2005wa,Janik:2007wt,Gromov:2008ie} to look at corrections
to giant magnons \cite{Hofman:2006xt}. These initially have $E=\Delta-J/2$ finite,
with $\Delta$ infinite, and we will be interested in exponential corrections starting
with these:
\begin{equation}
\delta E^{F}=\underset{\vphantom{1^{\frac{1}{1^{1}}^{1}}}\mbox{ first F-term }}{e^{-\Delta/2h}a_{1,0}}+\underset{\vphantom{1^{\frac{1}{1^{1}}^{1}}}\mbox{ second F-term }}{\big(e^{-\Delta/2h}\big)^{2}a_{2,0}}+\:\mathcal{O}\big(e^{-\Delta/2h}\big)^{3}.\label{eq:intro-F1-F2}
\end{equation}
We will calculate these for the correspondence between ABJM theory and IIA strings
in $AdS_{4}\times CP^{3}$ \cite{Aharony:2008ug,Klose:2010ki}, using the all-loop
S-matrix of \cite{Ahn:2008aa}. (For the related asymptotic Bethe ansatz, see \cite{Gromov:2008qe}.)
L\"{u}scher corrections in this theory were also studied by \cite{Bombardelli:2008qd,Lukowski:2008eq,Ahn:2008wd,Ahn:2010eg}.

Our first goal is to compare these F-terms to the string theory calculation of \cite{Abbott:2010yb}.
There we computed semiclassical corrections to the energy of giant magnons, both
in infinite volume, and obtaining a series of exponential corrections. The subtleties
of that calculation arose from the presence of heavy modes, and how to impose a cutoff
on these. The first F-term depends only on the light modes, but the second F-term
also has a contribution from the heavy modes. 

The existence of heavy and light modes is a novel feature of this version of the
correspondence, compared to the more familiar SYM / $AdS_{5}\times S^{5}$ case.
From the sigma-model perspective they are simply half the directions in target space,
where some have radius of curvature $R$ and some $R/2$. (Although some interesting
effects are seen at one loop, see \cite{Zarembo:2009au,Abbott:2011xp}.) By contrast
the spin chain has only the light degrees of freedom, and there are no physical bound
states corresponding to the heavy modes visible in the S-matrix. 

There are however bound states corresponding to particles in the mirror theory, which
is where the particle running in the loop lives \cite{Arutyunov:2007tc,Bajnok:2010ke}.
By including these when calculating the second F-term, we are able to match the term
coming from the heavy modes. For this, we need the bound-state S-matrix $S_{2-1}$
derived by \cite{Arutyunov:2008zt,Bajnok:2008bm}. At the same order there is also
a contribution from the light mode circling the space twice, of the type studied
by \cite{Heller:2008at}. Further terms of this form (including heavy modes circling
the space several times) allow us to get all orders of F-terms.

Dyonic giant magnons are of course physical bound states (of $Q\sim h$ of the same
type of particle). These have been studied by \cite{Chen:2006gq,Hatsuda:2008gd,Hatsuda:2009pc},
and the appropriate S-matrix is simply constructed by fusion: $S_{1-Q}=\prod^{Q}S_{1-1}$.
We are able to extend our calculation of second (and higher) order F-terms to the
dyonic case by similarly deriving a mixed S-matrix $S_{2-Q}=\prod^{Q}S_{2-1}$ for
virtual--physical scattering. 

In addition to F-terms, which are integrals over Euclidean momentum, poles in the
S-matrix give rise to $\mu$-terms. Because the pole is not at the saddle point of
the integral, these come with a different exponential factor. We will calculate the
following:
\begin{equation}
\delta E^{\mu}=e^{-\Delta/E}\Big[h\: a_{0,1}^{\mathrm{class.}}+a_{0,1}^{\mathrm{subl.}}+\mathcal{O}\Big(\frac{1}{h}\Big)\Big].\label{eq:intro-mu}
\end{equation}
The leading $\mu$-term gives rise to the classical (order $h\sim\sqrt{\lambda}$)
finite-volume correction, as first studied by \cite{Arutyunov:2006gs}. Here we will
be able to recover the result of \cite{Abbott:2009um}, for an elementary dyonic
giant magnon (in $CP^{2}$). The analogous agreement in $AdS_{5}\times S^{5}$ is
between \cite{Hatsuda:2008gd} and \cite{Minahan:2008re}. We then extend this calculation
to give the subleading term, providing a one-loop prediction. Some similar terms
were calculated by \cite{Gromov:2008ec,Bombardelli:2008qd}. 

In addition to the elementary giant magnon and its dyonic version \cite{Abbott:2009um,Hollowood:2009sc},
there are various other giant magnons which are understood to be superpositions of
two of these \cite{Hollowood:2009sc,Hatsuda:2009pc}. We calculate similar F-terms
for all of these.

\subsection*{Outline}

We begin with the F-term calculations in section \ref{sec:Lscher-F-term-Corrections},
reviewing the calculation of the first F-term before turning to the second F-term,
and then to all higher orders. The $\mu$-term corrections arise from poles in the
same integrals, and we treat these in section \ref{sec:Lscher-mu-term-Corrections}.
We then turn to composite magnons ($RP^{3}$ and the `big magnon') in section \ref{sec:F-terms-for-Composite-Magnons}. 

In the appendix, we calculate $\mu$-terms for the magnon in $S^{5}$, review the
various two-particle and bound-state S-matrices, and finally write some formulae
for the algebraic curve.

\section{L\"{u}scher F-term Corrections\label{sec:Lscher-F-term-Corrections}}

The basic formula for the F-term is 
\begin{equation}
\delta E^{F}=\fint_{-\infty}^{\infty}\frac{dq}{2\pi}\:\left(1-\frac{\varepsilon'(p)}{\varepsilon'(q_{\star})}\right)e^{-iq_{\star}L}\sum_{b}(-1)^{F_{b}}\left[S_{ba}^{ba}(q_{\star},p)-1\vphantom{\frac{1}{1}}\right].\label{eq:dE-F-basic}
\end{equation}
This gives an energy correction to a particle of type $a$ (and momentum $p$) due
to a virtual particle of any type $b$ circling the cylinder, size $L$. This formula
was first derived for a relativistic system $\varepsilon(p)=\sqrt{p^{2}+m^{2}}$
\cite{Luscher:1985dn,Klassen:1990ub} but holds for an arbitrary dispersion relation
$\varepsilon(p)$ \cite{Janik:2007wt}.%
\footnote{Note that \cite{Janik:2007wt} are missing a factor $(-1)^{F}$ which was restored
by \cite{Gromov:2008ie} and \cite{Heller:2008at}. Note also that the virtual particle
need not have the same dispersion relation as the real particle; in general we may
sum over several kinds of them, each with some $\varepsilon_{b}(q_{\star})$.%
} The momentum $q_{\star}$ is defined as a function of $q$ by the on-shell condition
\begin{equation}
q^{2}+\varepsilon^{2}(q_{\star})=0\,.\label{eq:q-qstar-equation}
\end{equation}
The integration contour in \eqref{eq:dE-F-basic} has the Euclidean energy $q$ real,
and thus $q_{\star}$ is imaginary. In this notation the \emph{Lorentzian} two-momenta
of the real and virtual particles are 
\[
p_{\mu}=(\epsilon(p),p)\qquad\mbox{and}\qquad q_{\mu}=(iq,q_{\star})\,.
\]
It is the fact that both particles are on-shell which allows one to replace the (infinite-volume)
four-point vertex $G_{4}(-p_{\mu},-q_{\mu},p_{\mu},q_{\mu},)$ with the asymptotic
S-matrix $S(q_{\star},p)$ when deriving \eqref{eq:dE-F-basic}. This is a consequence
of moving the $\int dq_{1}$ contour and crossing a pole of the propagator $G(q_{\mu})$
located at $q_{1}=q_{\star}$, while the other component of the loop integration
$\int dq_{0E}$ survives in the resulting formula. 

While this formula does not assume integrability, it has often been useful there,
since the S-matrix plays such a central role. In the AdS/CFT context, these corrections
were calculated for giant magnons in $S^{5}$ by \cite{Janik:2007wt,Heller:2008at,Gromov:2008ie},
and in $CP^{3}$ by \cite{Bombardelli:2008qd,Lukowski:2008eq}. They provide an important
check on the large-volume expansion of the thermodynamic Bethe ansatz (TBA) equations. 

The dispersion relation for the $CP^{3}$ giant magnon we are interested in is 
\begin{equation}
\mathcal{E}_{Q}(p)=\sqrt{\frac{Q^{2}}{4}+4\, h^{2}\:\sin^{2}\frac{p}{2}\,}\,.\label{eq:disp-rel-Q-1}
\end{equation}
Here $Q=1$ is the case of a single elementary magnon, for which we write $\varepsilon(p)=2h\,\sin\frac{p}{2}+\mathcal{O}(1/h)$,
and $h=\sqrt{\lambda/2}+\mathcal{O}(\lambda^{0})$ is the coupling. 

We study generalisations of existing calculations in two directions: 
\begin{itemize}
\item To higher-order F-terms, in which the virtual particle either circles the cylinder
more than once, or else is replaced by a bound state. 
\item To treat dyonic magnons $Q\sim h$, for both first and higher F-terms. 
\end{itemize}
In both cases the corresponding string calculations were given in \cite{Abbott:2010yb},
using the algebraic curve. There one obtains an integral very similar to \eqref{eq:dE-F-basic},
and it is easiest to simply compare the integrands.

\subsection{Review of the simplest comparison}

Before we get started with generalisations, let us recall how to connect the F-term
formula \eqref{eq:dE-F-basic} to algebraic curve results, and fix some notation. 

The real particle $a=1$ with two-momentum $(\varepsilon(p),p)$ is described using
the Zhukovski variables $x^{\pm}$ as usual, see \eqref{eq:defn-Xp-Xm-Qp}, and for
$Q=1$ these obey
\begin{equation}
x^{\pm}=e^{\pm ip/2}+\mathcal{O}(\frac{1}{h}).\label{eq:xp-xm-nondyonic}
\end{equation}
For the virtual particle $b$ with $(iq,q_{\star})$, we call the spectral parameters
$y^{\pm}$. It is useful to define $x$ by \cite{Gromov:2008ie,Bombardelli:2008qd}
\begin{equation}
x+\frac{1}{x}=y^{\pm}+\frac{1}{y^{\pm}}\pm\frac{1}{2ih}\label{eq:simplest-defn-x}
\end{equation}
which implies $Q(y^{\pm})=1$ exactly, and the following expansions:
\begin{align}
y^{\pm} & =x\pm\frac{i\, x^{2}}{2h(x^{2}-1)}+\mathcal{O}\Big(\frac{1}{h^{2}}\Big)\nonumber \\
q_{\star} & =-i\,\log\frac{y^{+}}{y^{-}}=\frac{1}{h}\frac{x}{x^{2}-1}+\mathcal{O}\Big(\frac{1}{h^{3}}\Big)\label{eq:simplest-y-qstar-q}\\
q & =-i\,\varepsilon(q_{\star})=\frac{i}{2}\left(\frac{x^{2}+1}{x^{2}-1}\right)+\mathcal{O}\Big(\frac{1}{h^{2}}\Big).\nonumber 
\end{align}
From the last equation we see that the integral along the real line of $q$ will
become an integral along the upper half unit circle in $x$. And on this circle,
$q_{\star}$ is imaginary, as well as small. We will also need: 
\begin{equation}
\varepsilon'(q_{\star})=-i\, h\frac{2x}{x^{2}+1}+\mathcal{O}\Big(\frac{1}{h}\Big)\label{eq:simplest-epsilonprime}
\end{equation}

The S-matrix of \cite{Ahn:2008aa} for the scattering of a particle type A (and $a=1$)
with one of type A or B (and $b=1,2,3,4$) can be written as follows:
\begin{align*}
S_{b}(y^{\pm},x^{\pm}) & \;=S^{AA}\;=\hat{S}_{b1}^{b1}\: n\,\sigma\\
\tilde{S}_{b}(y^{\pm},x^{\pm}) & \;=S^{AB}\;=\hat{S}_{b1}^{b1}\:\tilde{n}\,\sigma\,.
\end{align*}
Here our notation is to write 
\[
n(y^{\pm},x^{\pm})=\frac{1-\frac{1}{y^{+}x^{-}}}{1-\frac{1}{y^{-}x^{+}}},\qquad\tilde{n}(y^{\pm},x^{\pm})=\frac{y^{-}-x^{+}}{y^{+}-x^{-}}\,.
\]
As usual $\sigma$ is the BES dressing phase \cite{Beisert:2006ez,Vieira:2010kb},
and the relevant terms of the $su(2|2)$ invariant matrix part \cite{Beisert:2005tm,Arutyunov:2006yd}
are: 
\[
\hat{S}=a_{1}\, E_{11}^{11}+(a_{1}+a_{2})E_{21}^{21}+a_{6}(E_{31}^{31}+E_{41}^{41})
\]
where, in the string frame, 
\begin{align*}
a_{1}(y^{\pm},x^{\pm}) & =\frac{1}{\tilde{n}(y^{\pm},x^{\pm})}\;\sqrt{\frac{x^{+}}{x^{-}}}\:\sqrt{\frac{y^{-}}{y^{+}}}\,.
\end{align*}
We give all the other components in appendix \ref{sub:Two-particle-S-matrix}. Using
the $y^{\pm}=x+\mathcal{O}(1/h)$ as above, they become simply 
\begin{equation}
a_{1}=\frac{x-x^{-}}{x-x^{+}}\sqrt{\frac{x^{+}}{x^{-}}}+\mathcal{O}\Big(\frac{1}{h}\Big),\qquad a_{2}=0+\ldots,\qquad a_{6}=1+\ldots\label{eq:simplest-a1a2a6}
\end{equation}
and the phase parts become 
\begin{gather}
\sigma=\sigma_{\mathrm{AFS}}+\mathcal{O}\Big(\frac{1}{h}\Big)=\frac{x-1/x^{+}}{x-1/x^{-}}e^{-\frac{i\, x}{h(x^{2}-1)}\left(E-\frac{Q}{2}\right)}+\mathcal{O}\Big(\frac{1}{h}\Big)\label{eq:simplest-sigma-n}\\
n=\frac{x-1/x^{-}}{x-1/x^{+}}+\ldots,\qquad\tilde{n}=\frac{x-x^{+}}{x-x^{-}}+\ldots\,.\nonumber 
\end{gather}

We are now ready to compare to the algebraic curve calculation. The kinematic factor
above gives essentially the frequency factor of the integral there:%
\footnote{More correctly, on the right hand side we had in \cite{Abbott:2010yb} 
\[
\frac{1}{4\pi}\fint_{\mathbb{U}_{+}}dx\,\partial_{x}\Omega_{45}(x)\;\ldots+\frac{1}{4\pi}\fint_{\mathbb{U}_{-}}dx\,\partial_{x}\Omega_{45}(x)\;\ldots
\]
Both halves of this integral give the same contribution to the F-term. %
}
\begin{equation}
\fint_{\mathbb{R}}\frac{dq}{2\pi i}\,\left(1-\frac{\varepsilon'(p)}{\varepsilon'(q_{\star})}\right)\;\ldots=\frac{1}{2\pi}\int_{\mathbb{U}_{+}}dx\,\partial_{x}\Omega_{45}(x)\;\ldots\,.\label{eq:simplest-kinematic}
\end{equation}
The remaining factor in the integrand is 
\begin{align*}
e^{-iq_{\star}L}\sum_{b}(-1)^{F_{b}}\left(S_{b}+\tilde{S}_{b}\right)(y^{\pm},x^{\pm}) & =e^{-iq_{\star}L}\left[\vphantom{1_{1}^{1^{1}}}a_{1}+(a_{1}+a_{2})-2a_{6}\right]\sigma(n+\tilde{n})\\
 & =\left[2a_{1}(x,x^{\pm})-2\right]2e^{-\frac{i\, x}{h(x^{2}-1)}\left(L+E-\frac{Q}{2}\right)}+\mathcal{O}\Big(\frac{1}{h}\Big)\\
 & =F_{\mathrm{light}}^{(\ell=1)}\Big\vert_{Q=1}+\mathcal{O}\Big(\frac{1}{h}\Big).
\end{align*}
The notation from \cite{Abbott:2010yb} which we use here is that 
\begin{align}
\Omega_{45}(x) & =\frac{1}{x^{2}-1}\left(1-x\frac{X^{+}+X^{-}}{X^{+}X^{-}+1}\right)\label{eq:Ohm45}\\
F_{\mathrm{light}}^{(\ell)} & =\sum_{ij\;\mathrm{light}}(-1)^{F_{ij}}e^{-i\,\ell\left[q_{i}(x)-q_{j}(x)\right]}.\nonumber 
\end{align}
The off-shell frequency $\Omega_{ij}(x)$ is the same for all light modes, and twice
this for all heavy modes. In the exponent $q_{i}(x)$ are the quasimomenta, which
have the following poles at $x=1$ \cite{Gromov:2008bz}: 
\begin{equation}
q_{i}(x)=\begin{cases}
\alpha\frac{x}{x^{2}-1}+\mathcal{O}(x-1)^{0}, & i=1,2,3,4\\
\;0+\mathcal{O}(x-1)^{0}, & i=5,6\\
-\alpha\frac{x}{x^{2}-1}+\mathcal{O}(x-1)^{0}, & i=7,8,9,10,\qquad\mbox{where }\alpha=\Delta/h\,.
\end{cases}\label{eq:poles-of-quasimom}
\end{equation}
Light modes are polarisations $(i,j)$ in which one of the sheets is $5$ or $6$,
while heavy modes connect one sheet $i\leq4$ to another with $j\geq7$. Thus the
contribution $F_{\mathrm{light}}$ comes with $\exp(-\frac{i}{h}\frac{x}{x^{2}-1}\Delta)=\exp(-\frac{i}{h}\frac{x}{x^{2}-1}(E+J/2))$,
and so for the agreement of exponents above, we have set 
\begin{equation}
L=\frac{J}{2}\label{eq:L-equals-J/2}
\end{equation}
and dropped the $Q/2$ appearing in \eqref{eq:simplest-sigma-n}, since this is order
$1$. 

\begin{figure}
\begin{centering}
\includegraphics[width=7cm]{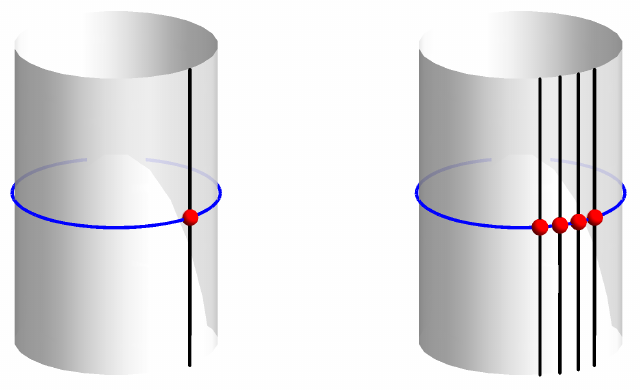}
\par\end{centering}

\caption{Sketch showing the first F-term, for a single magnon (left) and for a dyonic magnon
(right). In the dyonic case (drawn as if $Q=4$ although in reality $Q\sim h\gg1$)
we attempt to show that the relevant S-matrix is constructed by fusion: $S_{1-Q}=\prod^{Q}S_{1-1}$.
\label{fig:first-F-term}}
\end{figure}

\subsection{Dyonic magnon F-term}

The first step we take is to consider corrections to a dyonic magnon. This is a bound
state of $Q\sim h\gg1$ of the elementary particles of the same kind (taken to be
of type A, and $a=1$), and it is the pole in $a_{1}$ at $y^{-}=x^{+}$ which allows
the bound state to form. The constituent particles then have 
\[
x_{k}^{-}=x_{k-1}^{+},\qquad k=1,2,\ldots Q\,.
\]
The relevant bound-state S-matrix is simply the product of the constituent S-matrices,
which we illustrate as disjoint red balls in figure \ref{fig:first-F-term}. It has
the property that  all dependence on the intermediate $x_{k}^{+}$ cancels out, leaving
only $X^{-}=x_{1}^{-}$ and $X^{+}=x_{Q}^{+}$. (We review this construction in appendix
\ref{sub:Physical-bound-states-S}.) The result is
\begin{align*}
S_{b}(y^{\pm},X^{\pm})\equiv S_{1-Q}^{AA}(y^{\pm},X^{\pm})_{b1}^{b1} & =\prod_{k=1}^{Q}S_{b}(y^{\pm},x_{k}^{\pm})\\
 & =s_{b}(y^{\pm},X^{\pm})\: S_{\mathrm{BDS}}(y^{\pm},X^{\pm})\:\sqrt{\frac{X^{+}}{X^{-}}}\left(\frac{y^{-}}{y^{+}}\right)^{Q/2}\sigma(y^{\pm},X^{\pm})
\end{align*}
or, if the virtual particle is type B, 
\[
\tilde{S}_{b}(y^{\pm},X^{\pm})\equiv S_{1-Q}^{BA}(y^{\pm},X^{\pm})_{b1}^{b1}=s_{b}(y^{\pm},X^{\pm})\;\sqrt{\frac{X^{+}}{X^{-}}}\left(\frac{y^{-}}{y^{+}}\right)^{Q/2}\sigma(y^{\pm},X^{\pm})\,.\hphantom{\: S_{\mathrm{BDS}}(y^{\pm},X^{\pm})}
\]
Then we want to evaluate the following sum: 
\begin{align}
e^{-iq_{\star}L}\sum_{b}(-1)^{F_{b}}(S_{b}+\tilde{S}_{b}) & =e^{-iq_{\star}L}\left(S_{\mathrm{BDS}}+1\right)\left(s_{1}+s_{2}-s_{3}-s_{4}\right)\sqrt{\frac{X^{+}}{X^{-}}}\left(\frac{y^{-}}{y^{+}}\right)^{Q/2}\sigma(y^{\pm},X^{\pm})\label{eq:sum-S-dyonic-F1}\\
 & =e^{-iq_{\star}L}\left(\frac{n^{2}}{\tilde{n}^{2}}+1\right)\left(1+\frac{\tilde{n}}{n}-2\tilde{n}\: e^{-ip/2}\right)e^{ip/2}\frac{1}{n}e^{-\frac{i}{h}\frac{x}{x^{2}-1}(E-Q/2)}+\mathcal{O}\Big(\frac{1}{h}\Big)\nonumber \\
 & =F_{\mathrm{light}}^{(\ell=1)}+\mathcal{O}\Big(\frac{1}{h}\Big)\nonumber 
\end{align}
where we have used $y^{\pm}=x+\mathcal{O}(1/h)$ and that \eqref{eq:simplest-sigma-n}
still holds in the dyonic case. 

This is the dyonic generalisation of the term calculated in \cite{Bombardelli:2008qd},
and the same result was also found by \cite{Ahn:2010eg}. It matches the algebraic
curve calculation we gave in \cite{Abbott:2010yb}.

\subsection{Higher F-terms\label{sub:Higher-F-terms-light}}

By higher F-terms we mean those suppressed by%
\footnote{These we referred to as `sub-subleading' corrections in the algebraic curve calculation
\cite{Abbott:2010yb}, in which the ordinary F-term was subleading to the infinite-volume
one-loop correction. But in this paper we want to distinguish these from subsequent
terms in $1/h$.%
} 
\[
\left(e^{-\Delta/2h}\right)^{m},\qquad m=2,3,4\ldots\,.
\]
Such L\"{u}scher terms were studied in \cite{Heller:2008at}, where they arose from
diagrams in which a virtual particle circles the cylinder twice (or $m$ times).
The calculation performed there is a semiclassical mode sum, in which the mode $b$
has a phase shift $e^{i\delta_{ba}(k,p)}=S_{ba}^{ba}(k,p)$. To get the $m=2$ term,
they use a cylinder of size $2L$ and a phase shift $2\delta$. Adding up all terms
leads them to the following formula (their equation (33) or (3.8)):
\begin{equation}
\delta E=\frac{-1}{2\pi i}\fint_{-\infty}^{\infty}dq_{\star}\,\left[\varepsilon'(q_{\star})-\varepsilon'(p)\right]\sum_{b}(-1)^{F_{b}}\log\left(\frac{1-S_{b1}^{b1}(q_{\star},p)e^{-iq_{\star L}}}{1-e^{-iq_{\star L}}}\right).\label{eq:HJL-all-orders}
\end{equation}
Expanding this with $-\log\left(\frac{1-S\, e^{-L}}{1-e^{-L}}\right)=e^{-L}(S-1)+\frac{1}{2}e^{-2L}(S^{2}-1)+\frac{1}{3}e^{-3L}(S^{3}-1)+\ldots$
we write the $m$ component as 
\begin{equation}
\delta E_{\mathrm{HJL}}^{F,m}=\frac{1}{2\pi}\fint_{-\infty}^{\infty}dq\,\left(1-\frac{\varepsilon'(p)}{\varepsilon'(q_{\star})}\right)e^{-miq_{\star}L}\frac{1}{m}\sum_{b}(-1)^{F_{b}}\left[S_{b1}^{b1}(q_{\star},p)^{m}-1\right].\label{eq:dE-HJL-m}
\end{equation}
Here we changed the integration variable, and also the contour --- \cite{Heller:2008at}'s
derivation has $q_{\star}$ real, and we assume that this can be rotated back to
imaginary in the same way as the $m=1$ term. Notice that we can drop the final $-1$
from this formula, since $\sum_{b}(-1)^{F_{b}}1=0$ as we have always equally many
bosons and fermions.

To apply this result to $CP^{3}$ giant magnons, we should extend the sum over $b$
to also include whether the virtual particle is of type A or B. Then we obtain the
following result, corresponding to the left half of figure \ref{fig:second-F-term}:
\begin{align}
e^{-2iq_{\star}L}\sum_{b=1}^{4}(-1)^{F_{b}}\left[(S_{b})^{2}+(\tilde{S}_{b})^{2}\right] & =e^{-2iq_{\star}L}\left[a_{1}^{2}+(a_{1}+a_{2})^{2}-2(a_{6})^{2}\right]\left(n^{2}+\tilde{n}^{2}\right)(\sigma_{\mathrm{AFS}})^{2}\nonumber \\
 & =e^{-2i\frac{\alpha\, x}{x^{2}-1}}4\left[a_{1}(x,x^{\pm})-1\right]+\mathcal{O}\Big(\frac{1}{h}\Big).\label{eq:sum-S-like-HJL}
\end{align}
The kinematic factor is the same as for the $m=1$ case, \eqref{eq:simplest-kinematic}. 

Now let us compare this to the algebraic curve calculation in \cite{Abbott:2010yb},
where we wrote:%
\footnote{This expansion in $\ell$ is also given in \cite{Gromov:2008ie}, their (4.1), but
missing the $1/\ell$ factor.%
} 
\begin{align}
\delta E_{\mathrm{new}}^{F} & =-\frac{1}{4i}\oint_{\mathbb{U}}dx\sum_{ij}(-1)^{F_{ij}}\frac{q'_{i}(x)-q'_{j}(x)}{2\pi}\cot\Big(\frac{q_{i}(x)-q_{j}(x)}{2}\Big)\:\Omega_{ij}(x)\label{dE-F-AC-with-cot}\\
 & =\sum_{\ell=1,2,3\ldots}\frac{-1}{4\pi i}\sum_{\pm}\fint_{\mathbb{U}_{\pm}}dx\sum_{ij}(-1)^{F_{ij}}e^{\mp\ell i(q_{i}-q_{j})}\frac{1}{\ell}\partial_{x}\Omega_{ij}(x).\label{eq:dE-F-AC-all-m}
\end{align}
The terms in this sum are not the same as those in \eqref{eq:dE-HJL-m}, because
the heavy modes at $\ell$ contribute to $m=2\ell$. (This is because both $q_{i}$
and $q_{j}$ contribute to the pole at $x=\pm1$, see \eqref{eq:poles-of-quasimom},
rather than just one for the light mode.) Thus the $m=1$ term checked in \cite{Abbott:2010yb,Ahn:2010eg}
involves only the light modes, while the $m=2$ term involves also the heavy modes.
We wrote this term as 
\begin{equation}
\delta E^{F,2}=-\frac{1}{2\pi i}\fint_{\mathbb{U}_{+}}dx\left[\frac{1}{2}\; F_{\mathrm{light}}^{(\ell=2)}\;\Omega_{45}'(x)+F_{\mathrm{heavy}}^{(\ell=1)}\;2\Omega_{45}'(x)\right]\label{eq:dE-F-AC-m2}
\end{equation}
and had the following expressions for the $\sum_{ij}(-1)^{F_{ij}}e^{-\ell i(q_{i}-q_{j})}$
factor: 
\begin{align}
F_{\mathrm{light}}^{(\ell=2)}\Big\vert_{Q=1} & =e^{-2i\frac{\alpha x}{x^{2}-1}}\frac{4(x^{2}-1)\left(e^{ip}-1\right)}{(x-e^{ip/2})^{2}}\nonumber \\
F_{\mathrm{heavy}}^{(\ell=1)}\Big\vert_{Q=1} & =e^{-2i\frac{\alpha x}{x^{2}-1}}\frac{(x+1)\left(e^{ip/2}-1\right)\left(e^{ip/2}(3+x)-(3x+1)\right)}{\left(x-e^{ip/2}\right)^{2}}\,.\label{eq:F+heavy}
\end{align}
It is easy to see that the term coming from the light modes alone matches the $m=2$
term from HJL, \eqref{eq:dE-HJL-m} above. (The factor $\frac{1}{2}$ in \eqref{eq:dE-F-AC-m2}
is the $\frac{1}{m}$ in \eqref{eq:dE-HJL-m} of course. See section \ref{sub:All-higer-F-terms}
for comparison at general $m$.) The term from the heavy modes requires a different
explanation.

\subsection{Heavy modes and virtual bound states}

\begin{figure}
\begin{centering}
\includegraphics[width=7cm]{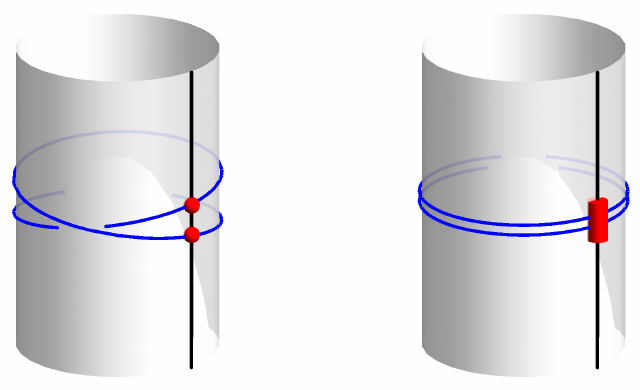}
\par\end{centering}

\caption{Sketch showing the second F-term for a single magnon. On the left is the contribution
of a virtual particle circling the space twice, as in HJL, producing $(S_{1-1})^{2}$.
On the right the contribution of a heavy mode (i.e. an $M=2$ bound state) for which
we need $S_{2-1}$. \label{fig:second-F-term}}
\end{figure}
The sum over virtual particles in \eqref{eq:dE-F-basic} should also run over all
possible bound states. However the bound states we are interested in are those in
the mirror theory, and because of this, we are interested in poles at $y^{+}=x^{-}$,
rather than those at $y^{-}=x^{+}$ needed to build the dyonic magnon above. 

In the ABJM case it is $\tilde{S}=S^{AB}$ which has these poles. We can read off
the list of possible AB bound states from \eqref{eq:full-S-hat} -- only $E_{11}^{11}$
and $E_{22}^{22}$ come with $a_{1}$ which cancels the pole from $\tilde{n}$. Thus
we obtain the following list: 
\begin{equation}
\begin{array}{rl}
\mbox{bose+bose:} & \mbox{1+2}\\
\mbox{fermi+fermi:} & \mbox{3+3},\;\mbox{3+4},\;\mbox{4+4}\vspace{2mm}\\
\mbox{bose+fermi:} & \mbox{1+3},\;\mbox{1+4},\;\mbox{2+3},\;\mbox{2+4}\,.
\end{array}\label{eq:heavy-list}
\end{equation}
This is very similar to the off-shell construction of heavy modes in the algebraic
curve, \cite{Bandres:2009kw,Abbott:2010yb}, where again there is one heavy boson
made from a pair of light bosons, and three made from a pair of fermions. 

The first rather naive way to proceed is to use for S-matrix simply the product of
those for the constituent light modes. This does lead us to the right answer: 
\begin{align}
e^{-2iq_{\star}L}\sum_{b+c=\mathrm{heavy}}(-1)^{F_{b+c}}S_{b}\tilde{S}_{c} & =\left[a_{1}(a_{1}+a_{2})+3(a_{6})^{2}-2a_{1}a_{6}-2(a_{1}+a_{2})a_{6}\vphantom{\frac{1}{1}}\right]n\tilde{n}\:\sigma^{2}\label{eq:heavy-sumS-adhoc}\\
 & =e^{-2i\frac{\alpha\, x}{x^{2}-1}}(a_{1}^{2}-4a_{1}+3)+\mathcal{O}\Big(\frac{1}{h}\Big)\nonumber \\
 & =F_{\mathrm{heavy}}^{(\ell=1)}\Big\vert_{Q=1}+\ldots\,.\nonumber 
\end{align}
We have used here that the parameters of the constituents are all $y_{k}^{\pm}=x+\mathcal{O}(1/h)$,
and thus $a_{i}$ simplify exactly as before, \eqref{eq:simplest-a1a2a6}. In full,
defining $x$ by \eqref{eq:simplest-defn-x} with $\pm M/2ih$ on the right, and
taking $M=2$, the constituent points are%
\footnote{Note that despite this virtual bound state arising from a different pole, we can
still overlap these $y_{k}^{-}=y_{k-1}^{+}$ as for the real bound state. This is
simply a choice of how to label them. %
}
\begin{equation}
\begin{array}{ccccccccl}
Y^{-} & \negthickspace\negthickspace=\negthickspace\negthickspace & y_{1}^{-} &  &  &  &  & =\negthickspace\negthickspace & x-\frac{ix^{2}}{h(x^{2}-1)}+\mathcal{O}(\frac{1}{h^{2}})\\
 &  & y_{1}^{+} & \negthickspace\negthickspace=\negthickspace\negthickspace & y_{2}^{-} &  &  & =\negthickspace\negthickspace & x\\
 &  &  &  & y_{2}^{+} & \negthickspace\negthickspace=\negthickspace\negthickspace & Y^{+} & =\negthickspace\negthickspace & x+\frac{ix^{2}}{h(x^{2}-1)}+\ldots\,.
\end{array}\label{eq:evalF2}
\end{equation}
Using these the AFS dressing phase is the square of that in \eqref{eq:simplest-sigma-n}:
\[
\sigma(y_{1}^{\pm},x^{\pm})\sigma(y_{2}^{\pm},x^{\pm})=\sigma(Y^{\pm},x^{\pm})=\left(\frac{x-1/x^{+}}{x-1/x^{-}}\right)^{2}e^{-\frac{2i}{h}\frac{x}{x^{2}-1}\left(E-\frac{Q}{2}\right)}+\mathcal{O}\Big(\frac{1}{h}\Big).
\]

Less naively, we should use the bound-state S-matrix $S_{2-1}(Y^{\pm},x^{\pm})_{b1}^{b1}$
derived by \cite{Arutyunov:2008zt,Bajnok:2008bm}. The particular matrix elements
we need are the same ones recently used in \cite{Ahn:2010yv} (although here we turn
off the $\beta$ twist factors), which correspond to the transfer matrix eigenvalues
used in \cite{Gromov:2009tv,Serban:2010sr}:
\[
T_{M,1}^{SU(2)}=(-1)^{M}\left[(M+1)\, a_{5}^{5}+(M-1)\:2a_{8}^{8}-M\: a_{9}^{9}-M\,\frac{a_{9}^{9}+a_{3}^{3}}{2}\right].
\]
 Using the appropriate prefactors for ABJM (as reviewed in appendix \ref{sub:Bound-states-in-ABBN}),
we are led to the following expression, corresponding to the right half of figure
\ref{fig:second-F-term}:
\begin{align*}
e^{-2iq_{\star}L}\sum_{b=1}^{8}(-1)^{F_{i}^{i}}S_{b} & =e^{-2iq_{\star}L}\left[\vphantom{\frac{1}{1}}3a_{5}^{5}+2a_{8}^{8}-2a_{9}^{9}-(a_{9}^{9}+a_{3}^{3})\right](x^{\pm},Y^{\pm})\\
 & \qquad\times n(y_{1}^{\pm},x^{\pm})\tilde{n}(y_{2}^{\pm},x^{\pm})\;\sigma(Y^{\pm},x^{\pm})\\
 & =F_{\mathrm{heavy}}^{(\ell=1)}\Big\vert_{Q=1}+\mathcal{O}\Big(\frac{1}{h}\Big).
\end{align*}

Note that while we are using the correct mirror--physical bound-state matrix elements
here, we are still using a naive analytic continuation of the physical dressing phase
$\sigma(x^{\pm},Y^{\pm})$ in order to evaluate it at this $Y^{\pm}$. This is commonly
done for F-term calculations, as for instance in \cite{Bombardelli:2008qd,Lukowski:2008eq,Ahn:2008wd,Ahn:2010eg},
and it does give the correct answer here. We have also checked that the strong-coupling
limit of the mirror\textendash{}physical dressing phase computed in \cite{Arutyunov:2009kf}
reduces to this.

\subsubsection*{Kinetic factor }

In \eqref{eq:dE-F-basic} we are allowed to use two different dispersion relations
for the real and the virtual particles. Since $y_{k}^{\pm}$ form a bound state,
we should use for the virtual particle $\mathcal{E}(q_{\star})=\mathcal{E}_{M}(q_{\star})$.
For clarity let us write this for general $M$, where we will have $Y^{\pm}=x\pm\frac{i\, M\, x^{2}}{2h(x^{2}-1)}+\mathcal{O}(\frac{1}{h^{2}})$
and thus
\[
q_{\star}=\frac{M}{h}\frac{x}{x^{2}-1}+\mathcal{O}\Big(\frac{1}{h^{3}}\Big),\qquad q=\frac{i\, M}{2}\left(\frac{x^{2}+1}{x^{2}-1}\right)+\mathcal{O}\Big(\frac{1}{h^{2}}\Big).
\]
Then \eqref{eq:simplest-epsilonprime} is unchanged at this order: 
\[
\mathcal{E}'(q_{\star})=h\frac{2x}{x^{2}+1}+\mathcal{O}\Big(\frac{1}{h}\Big).
\]
Define $\bar{q}_{\star}=q_{\star}/M$ and $\bar{q}=q/M$ , which are essentially
the momentum and energy of each of the constituents of the virtual bound state. Then
in terms of these, the on-shell condition can be re-written
\[
q^{2}+\mathcal{E}(q_{\star})^{2}=0\quad\Rightarrow\quad\bar{q}^{2}+\varepsilon(\bar{q}_{\star})^{2}=0+\mathcal{O}\Big(\frac{M}{h^{2}}\Big).
\]
and thus $(i\bar{q},\bar{q}_{\star})$ describe a single on-shell particle, \eqref{eq:q-qstar-equation}.
We can now re-write the natural kinetic factor from \eqref{eq:dE-F-basic} in terms
of the new barred variables:

\begin{equation}
\fint dq\left(1-\frac{\epsilon'(p)}{\mathcal{E}'(q_{\star})}\right)e^{-i\, q_{\star}\, L}\ldots=\fint d\bar{q}\left(1-\frac{\epsilon'(p)}{\varepsilon'(\bar{q}_{\star})}\right)e^{-i\, M\bar{q}_{\star}\, L}\: M\times\ldots\,.\label{eq:heavy-kinematic}
\end{equation}
Now since $\bar{q}=\frac{1}{h}\frac{x}{x^{2}-1}$, we can drop the bar and combine
this with the light mode integral, and then write this in terms of $x$ as before.
The only new features are an overall factor of $M$, and $ML$ in the exponent, both
of which are exactly what we wanted in order to match \eqref{eq:dE-F-AC-m2} when
$M=2$. 

Note, aside, that what we have written here would be equally true if we used for
the virtual particle instead $\mathcal{E}(q_{\star})=M\varepsilon(q_{\star}/M)$.
This is simply a superposition of two particles, like the $RP^{2}$ giant magnon.

\subsection{Dyonic second F-term\label{sub:Dyonic-second-F-term}}

\begin{figure}
\begin{centering}
\includegraphics[width=7cm]{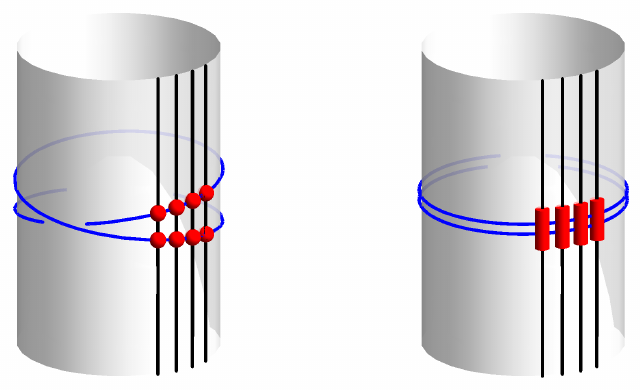}
\par\end{centering}

\caption{Sketch showing the second F-term for a dyonic magnon. For the light mode running
in the loop (left) we need $(S_{1-Q})^{2}$. For a heavy mode running in the loop
(right) we need $S_{2-Q}$, which we construct by fusion as $\prod^{Q}S_{2-1}$.\label{fig:second-F-term-Dyonic}}
\end{figure}
We can now extend this to allow the external particle to be a physical bound state,
i.e. a dyonic giant magnon. We will still have the same two contributions, this time
illustrated in figure \ref{fig:second-F-term-Dyonic}.  The first is from a fundamental
virtual particle going around twice: 
\begin{align*}
 & e^{-2iq_{\star}L}\sum_{b}(-1)^{F_{b}}\left[S_{b}(y^{\pm},X^{\pm})^{2}+\tilde{S}_{b}(y^{\pm},X^{\pm})^{2}\right]\\
 & =e^{-2iq_{\star}L}\left[(S_{\mathrm{BDS}})^{2}+1\vphantom{1^{1^{1}}}\right]\left[\vphantom{\frac{1}{1}}(s_{1})^{2}+(s_{2})^{2}-(s_{3})^{2}-(s_{4})^{2}\right]\frac{X^{+}}{X^{-}}\left(\frac{y^{-}}{y^{+}}\right)^{Q}\,\,\sigma^{2}(y^{\pm},X^{\pm})\\
 & =F_{\mathrm{light}}^{(\ell=2)}+\mathcal{O}\Big(\frac{1}{h}\Big).
\end{align*}
And the second contribution is from a virtual bound state running in the loop. For
this last contribution, we need to derive $S_{2-Q}$ from $S_{2-1}$ by fusion. We
give details of this in appendix \ref{sub:Constructing-S_2Q-by-fusion}, and here
simply write down the result: 
\begin{align}
e^{-2iq_{\star}L}\sum_{b=1}^{8}(-1)^{F}S_{b}(Y^{\pm},X^{\pm}) & =T_{0}(y_{1}^{\pm},y_{2}^{\pm},X^{\pm})\left[3t_{1}+t_{4}-2t_{5}-2t_{7}\vphantom{1^{1^{1}}}\right]\sigma(Y^{\pm},X^{\pm})\nonumber \\
 & =F_{\mathrm{heavy}}^{(\ell=1)}+\mathcal{O}\Big(\frac{1}{h}\Big).\label{eq:dyonic-F2-heavy}
\end{align}

We did not explicitly show the results of the corresponding algebraic curve calculation
in \cite{Abbott:2010yb} for the dyonic case. They do however match the present calculation
exactly. (We give all the necessary formulae in appendix \ref{sec:AC-Formulae}.)

\subsection{All higher F-terms\label{sub:All-higer-F-terms}}

It is possible to extend our results above to treat all higher F-terms. By doing
so we can recover exactly the corresponding algebraic curve results. We do this using
virtual elementary particles and two-particle ($M=2$) bound states. 

What is not entirely clear to us is how, without knowing about the string theory,
one would know to include $M=2$ but not for instance the $M=3,4$ virtual bound
states. There is no obvious distinction visible from the S-matrix. Certainly adding
such terms would spoil the agreement, starting with the 3rd F-term, because there
are no corresponding modes in the algebraic curve (or in the worldsheet sigma-model)
with mass 3 or 4.  In the analogous $AdS_{5}\times S^{5}$ calculation one does not
include any virtual bound states, and the analogous justification is to note that
all modes of the string have mass 1.

Let us now state the results. We obtain agreement of $\delta E^{F,m}$ for all $m$
as follows:
\begin{itemize}
\item For odd $m$ the only contribution is from the light modes. Comparing \eqref{eq:dE-HJL-m}
and \eqref{eq:dE-F-AC-all-m}, what we need is 
\[
e^{-imq_{\star}L}\sum_{b=1}^{4}(-1)^{F_{b}}\left[(S_{b})^{m}+(\tilde{S}_{b})^{m}\right]\;\:=\;\: F_{\mathrm{light}}^{(m)}=\negthickspace\sum_{ij\,\mathrm{light}}(-1)^{F_{ij}}e^{-im(q_{i}-q_{j})}+\mathcal{O}\Big(\frac{1}{h}\Big).
\]
This remains true if the physical particle is a dyonic magnon, interpreting $S_{b}$
as in \eqref{eq:sum-S-dyonic-F1}. 
\item For even $m$ there is in addition a contribution from the heavy modes circling the
space $\ell=m/2$ times. The agreement for this term comes from
\[
e^{-i\ell2q_{\star}L}\sum_{b=1}^{8}(-1)^{F_{b}}(S_{b})^{\ell}\;\:=\;\: F_{\mathrm{heavy}}^{(\ell)}=\negthickspace\sum_{ij\,\mathrm{heavy}}(-1)^{F_{ij}}e^{-i\ell(q_{i}-q_{j})}+\mathcal{O}\Big(\frac{1}{h}\Big).
\]
Here $e^{-im2q_{\star}L}$ is from \eqref{eq:heavy-kinematic}, appropriately generalised.
This agreement also holds for the dyonic case, where like \eqref{eq:dyonic-F2-heavy}
we use the mixed bound state S-matrix $S_{2-Q}$: 
\[
e^{-i\ell2q_{\star}L}T_{0}^{\ell}(y_{1}^{\pm},y_{2}^{\pm},X^{\pm})\left[3t_{1}^{\ell}+t_{4}^{\ell}-2t_{5}^{\ell}-2t_{7}^{\ell}\vphantom{1^{1^{1}}}\right]\sigma^{\ell}(Y^{\pm},X^{\pm})\;=\; F_{\mathrm{heavy}}^{(\ell)}+\mathcal{O}\Big(\frac{1}{h}\Big).
\]
We can write the total correction at this order as
\begin{equation}
\delta E^{F,m}=\frac{1}{2\pi}\fint_{-\infty}^{\infty}dq\,\left(1-\frac{\varepsilon'(p)}{\varepsilon'(q_{\star})}\right)\left[\frac{1}{m}F_{\mathrm{light}}^{(m)}+\frac{1}{m/2}2F_{\mathrm{heavy}}^{(m/2)}\right]+\mathcal{O}\Big(\frac{1}{h}\Big).\label{eq:dE-all-total}
\end{equation}

\end{itemize}
Both of these expansions in $m$ come from exact formulae, \eqref{eq:HJL-all-orders}
and \eqref{dE-F-AC-with-cot}. A convenient form in which to write the agreement
to all orders is the following:%
\footnote{Recall that $S_{b}=S_{1-Q}^{AA}(y^{\pm},X^{\pm})_{b1}^{b1}$, $\tilde{S}_{b}=S_{1-Q}^{AB}(y^{\pm},X^{\pm})_{b1}^{b1}$
in our notation, and $(S_{2-Q})_{b}=S(Y^{\pm},X^{\pm})_{b1}^{b1}$.%
} 
\begin{align}
\prod_{b=1}^{4}\left[\left(1-S_{b}e^{-iq_{*}L}\right)\left(1-\tilde{S}_{b}e^{-iq_{*}L}\right)\right]^{(-1)^{F_{b}}} & =\prod_{ij\,\mathrm{light}}\left[1-e^{-i(q_{i}-q_{j})}\right]^{(-1)^{F_{ij}}}+\mathcal{O}\Big(\frac{1}{h}\Big)\label{eq:all-equal-light}\\
\prod_{b=1}^{8}\left[1-(S_{2-Q})_{b}e^{-iq_{*}L}\right]^{(-1)^{F_{b}}} & =\prod_{ij\,\mathrm{heavy}}\left[1-e^{-i(q_{i}-q_{j})}\right]^{(-1)^{F_{ij}}}+\mathcal{O}\Big(\frac{1}{h}\Big).\label{eq:all-equal-heavy}
\end{align}

\section{L\"{u}scher $\mu$-term Corrections\label{sec:Lscher-mu-term-Corrections}}

In the field-theoretic derivation of the F-term formula \eqref{eq:dE-F-basic}, there
is one diagram whose $\int dq$ contour must be moved by $ip_{0}$ before combining
with the others. If this crosses any poles, they give rise to extra contributions,
which are called $\mu$-terms. The generic formula is
\begin{equation}
\delta E^{\mu}=\re\{\;-i\left(1-\frac{\varepsilon'(p)}{\varepsilon'(\tilde{q}_{\star})}\right)e^{-i\tilde{q}_{\star}L}\sum_{b}(-1)^{F_{b}}\res_{q=\tilde{q}}\left[S_{ba}^{ba}(q_{\star}(q),p)-1\vphantom{\frac{1}{1}}\right]\}\,.\label{eq:basic-mu-term}
\end{equation}
Not only are both real and virtual particles on-shell (as for the F-terms), but the
loop momentum $q_{\mu}$ is now completely fixed to discrete values: $\tilde{q}$
is Euclidean energy at which there is a pole, and $\tilde{q}_{\star}$ is the corresponding
momentum, given by \eqref{eq:q-qstar-equation}.

The first calculation of $\mu$-terms for giant magnons was \cite{Janik:2007wt},
recovering the classical finite-size corrections of \cite{Arutyunov:2006gs,Astolfi:2007uz}.
The dyonic version of this was studied by \cite{Hatsuda:2008gd}. 

For magnons in $AdS_{4}\times CP^{3}$, classical $\mu$-terms have been studied
by \cite{Lukowski:2008eq,Bombardelli:2008qd,Ahn:2008wd}. For the giant magnon in
$RP^{2}$, these papers correctly recovered the AFZ correction \cite{Grignani:2008te,Abbott:2009um}.
But for the elementary magnon (in $CP^{1}$) they obtained zero, apparently in contradiction
with the string sigma-model \cite{Lee:2008ui,Abbott:2008qd,Arutyunov:2006gs}. The
same zero result was also obtained by the first algebraic curve calculation \cite{Lukowski:2008eq},
\label{paragraph-non-d-mu-puzzles}and this was shown to be an order-of-limits problem
in \cite{Abbott:2009um}, but to date this has not been resolved in the literature
on L\"{u}scher corrections. 

We therefore compute $\mu$-terms for the \emph{dyonic} elementary magnon, i.e. for
the giant magnon solution in $CP^{2}$ of \cite{Abbott:2009um,Hollowood:2009sc}.
We are able to recover the classical algebraic curve result of \cite{Abbott:2009um}
for this correction. This classical term is the leading one in an expansion in $1/h$,
and we go on to evaluate the formula \eqref{eq:basic-mu-term} at the subleading
order. Such a subleading term was calculated by \cite{Gromov:2008ec} in $AdS_{5}\times S^{5}$
(where it was compared to a one-loop algebraic curve calculation) and by \cite{Bombardelli:2008qd}
for the case of the $RP^{2}$ giant magnon in $AdS_{4}\times CP^{3}$.

\subsection{Setup\label{sub:Setup-for-mu}}

The pole in the S-matrix which normally gives rise to \eqref{eq:basic-mu-term} is
at $y^{-}=X^{+}$. To find the pieces we need, solve the equation $Q(y^{\pm})=1$
for $y^{+}$, to get
\begin{equation}
y^{+}=X^{+}+\frac{i}{h}\frac{X^{+2}}{X^{+2}-1}+\frac{1}{h^{2}}\frac{X^{+3}}{(X^{+2}-1)^{3}}+\mathcal{O}\Big(\frac{1}{h^{3}}\Big).\label{eq:eval-mu1}
\end{equation}
Then the momentum of the virtual particle is 
\begin{align}
q_{\star}=-i\log\frac{y^{+}}{y^{-}} & =\frac{1}{h}\frac{X^{+}}{X^{+2}-1}-\frac{i}{2h^{2}}\frac{X^{+2}(X^{+2}+1)}{(X^{+2}-1)^{3}}+\mathcal{O}\Big(\frac{1}{h^{3}}\Big).\label{eq:setup-mu-qstar}
\end{align}
We will also need 
\begin{align*}
q=\pm i\varepsilon'(q_{\star}) & =\pm\frac{i}{2}\frac{X^{+2}+1}{X^{+2}-1}\mp\frac{1}{h^{2}}\frac{X^{+3}}{(X^{+2}-1)^{3}}+\mathcal{O}\Big(\frac{1}{h^{2}}\Big)\\
\varepsilon'(q_{\star}) & =2h\frac{X^{+}}{X^{+2}+1}-i\frac{X^{+2}}{(X^{+2}+1)^{2}}+\mathcal{O}\Big(\frac{1}{h}\Big)
\end{align*}
from which we get the kinetic factor to be 
\begin{equation}
\left(1-\frac{\mathcal{E}'_{Q}(p)}{\varepsilon'(q_{\star})}\right)=-\frac{(X^{+}-X^{-})(X^{+2}-1)}{2X^{+}(X^{+}X^{-}+1)}-\frac{i}{4h}\frac{\; X^{+}+X^{-}}{(X^{+}X^{-}+1)}+\mathcal{O}\Big(\frac{1}{h^{2}}\Big).\label{eq:setup-mu-kinetic}
\end{equation}
It is convenient to re-write the residue in terms of $y^{-}$ rather than $q$. Assuming
that we have only a simple pole, the Jacobian factor which this change inserts is
\begin{equation}
\lim_{q\to\tilde{q}}\frac{q-\tilde{q}}{y^{-}(q)-X^{+}}=\frac{1}{\partial_{q}y^{-}}=\mp2i\frac{X^{+}}{(X^{+2}-1)^{2}}\mp\frac{3}{h}\frac{X^{+2}(X^{+2}+1)}{(X^{+2}-1)^{4}}+\mathcal{O}\Big(\frac{1}{h^{2}}\Big).\label{eq:setup-mu-jacobian}
\end{equation}
Putting these pieces into \eqref{eq:basic-mu-term}, and keeping just the leading
term, we have:
\[
\delta E^{\mu}=\re\left\{ e^{-L\frac{iX^{+}}{h(X^{+2}-1)}}\frac{(X^{-}-X^{+})}{(X^{+2}-1)(X^{+}X^{-}+1)}\res_{Y^{-}=X^{+}}\sum_{b}(-1)^{F_{b}}S_{b}(y^{\pm},X^{\pm})\right\} .
\]

In the case we study here there will also be a pole at $y^{+}=X^{+}$. For this pole,
$y^{-}$ is given by 
\[
y^{-}=X^{+}-\frac{i}{h}\frac{X^{+2}}{X^{+2}-1}+\frac{1}{h^{2}}\frac{X^{+3}}{(X^{+2}-1)^{3}}+\mathcal{O}\Big(\frac{1}{h^{3}}\Big)
\]
and the only change to the formulae above is to change the sign of the last term
displayed for $q$, $q_{\star}$, the Jacobian and the kinetic factor. But $\varepsilon'(q_{\star})$
is unchanged.

\subsection{Leading $\mu$-term for the dyonic magnon\label{sub:Leading-mu-term}}

Using these pieces, we now wish to calculate \eqref{eq:basic-mu-term} considering
the real particle to be a dyonic giant magnon, and the virtual particle an elementary
magnon. As for the corresponding F-term calculation \eqref{eq:sum-S-dyonic-F1},
we are interested in 
\begin{equation}
\sum_{b}(-1)^{F_{b}}S_{b}(y^{\pm},X^{\pm})=(S_{\mathrm{BDS}}+1)\left[s_{1}+s_{2}-s_{3}-s_{4}\right]\sqrt{\frac{X^{+}}{X^{-}}}\left(\frac{y^{-}}{y^{+}}\right)^{Q/2}\sigma\,.\label{eq:sum-S-dyonic-mu}
\end{equation}
This has poles at $y^{\pm}=X^{+}$ arising only from $S_{\mathrm{BDS}}$, \eqref{eq:S-BDS}.
(Thus the only contribution is from when the virtual particle is type A.) We consider
first the pole at $y^{-}=X^{+}$. 
\begin{itemize}
\item The residue of $S_{\mathrm{BDS}}$, at point \eqref{eq:eval-mu1}, is: 
\[
\negthickspace\negthickspace\negthickspace\res_{y^{-}=X^{+}}S_{\mathrm{BDS}}=\frac{h(X^{+2}-1)}{iX^{+2}}\frac{(X^{+}-X^{-})^{2}(1-\frac{1}{X^{+}X^{-}})^{2}}{(1-\frac{1}{X^{+2}})^{2}}+\frac{X^{+}(X^{+}-X^{-})(X^{+}X^{-}-1)}{X^{-}(X^{+2}-1)}+\mathcal{O}\Big(\frac{1}{h}\Big).
\]
From the rest of $\hat{S}$ we need only $s_{1}=1$ at this order. 
\item The phases $\eta$ and the factor $e^{-iq_{\star}L}$ can be treated together \cite{Bombardelli:2008qd}:
\begin{align}
\negthickspace\negthickspace e^{-iq_{\star}L}\sqrt{\frac{X^{+}}{X^{-}}}\left(\frac{y^{-}}{y^{+}}\right)^{Q/2} & =e^{-iq_{\star}(L+\frac{Q}{2})}\sqrt{\frac{X^{+}}{X^{-}}}\label{eq:mu-expand-H}\\
 & =e^{-\frac{i}{h}(L+\frac{Q}{2})\frac{X^{+}}{X^{+2}-1}}\sqrt{\frac{X^{+}}{X^{-}}}\left[1+\left(L+\frac{Q}{2}\right)\frac{X^{+2}(X^{+2}+1)}{2h^{2}(X^{+2}-1)^{3}}+\mathcal{O}\Big(\frac{1}{h^{2}}\Big)\right].\nonumber 
\end{align}
Here we should note that this expansion appears to need to assume $L/h^{2}\ll1$.
(The same assumption was used in \cite{Bombardelli:2008qd}.) Of course for any L\"{u}scher
terms to be small corrections, we also need that $L/h\gg1$. Note however that the
term $L$ from here cancels out of the final correction $\delta E^{\mu}$, \eqref{eq:subleading-mu-final}
below. 
\item The AFS dressing phase is 
\begin{equation}
\negthickspace\negthickspace\sigma_{\mathrm{AFS}}=e^{\frac{-X^{+}}{X^{+2}-1}\left(\frac{1}{X^{-}}-\frac{1}{X^{+}}\right)}\left[\frac{X^{-}(X^{+2}-1)}{X^{+}(X^{+}X^{-}-1)}-i\frac{(X^{+}-X^{-})(X^{+2}-2X^{-}X^{+}+1)}{2h(X^{+2}-1)^{2}(X^{+}X^{-}-1)^{2}}+\mathcal{O}\Big(\frac{1}{h^{2}}\Big)\right].\label{eq:sigmaAFS}
\end{equation}

\end{itemize}
For the pole at $y^{+}=X^{+}$ the residue is exactly minus what we had above, and
all other factors the same (at this order). Following the contour prescription of
\cite{Hatsuda:2008gd}, this contribution enters with a minus, thus doubling the
result. Then using $L=J/2=\Delta-E$, the final result can be written 
\begin{align}
\delta E_{\mathrm{class.}} & =2\:\delta E_{\mathrm{class.}(y^{-}=X^{+})}\nonumber \\
 & =\mbox{\ensuremath{\mp}}2ih\: e^{-\frac{i\Delta}{h}\frac{X^{+}}{X^{+2}-1}}\sqrt{\frac{X^{-}}{X^{+}}}\frac{(X^{+}X^{-}-1)(X^{+}-X^{-})^{3}}{X^{-2}(X^{+}X^{-}+1)(X^{+2}-1)}\,.\label{eq:mu-leading-final}
\end{align}
This exactly matches the algebraic curve result in \cite{Abbott:2009um}, using $h=\sqrt{\lambda/2}=g/2$. 

The choice of sign here, which comes from $q=\pm i\varepsilon(q_{\star})$, matches
the factor $\cos(2\phi)$ seen in \cite{Abbott:2009um}. There $2\phi$ was interpreted
as the geometric angle between adjacent magnons, incorporating what \cite{Okamura:2006zv}
called type (i) and type (ii) solutions.

\subsection{Subleading $\mu$-term for the dyonic magnon\label{sub:Subleading-mu-term}}

In order to determine the subleading (in $1/h$) contributions to the $\mu$-term,
we already gave some expansions to more orders than required above. But there are
extra contributions that only appear to this order, which are:
\begin{itemize}
\item The remainder of $\hat{S}$ (i.e. the supertrace of the $su(2|2)$ bound-state S-matrix)
evaluated at \eqref{eq:eval-mu1} gives the following:
\begin{equation}
\left[s_{1}+s_{2}-s_{3}-s_{4}\right]=1+\frac{i}{h}\frac{X^{+}}{X^{+}-X^{-}}\left(\frac{X^{-}}{X^{+}X^{-}-1}-\frac{2X^{+}}{X^{+2}-1}\sqrt{\frac{X^{-}}{X^{+}}}\right)+\mathcal{O}\Big(\frac{1}{h^{2}}\Big).\label{eq:str}
\end{equation}
For the pole at $y^{+}=X^{+}$, this is instead exactly 1. 
\item The Hernandez--Lopez dressing phase \cite{Hernandez:2006tk} also contributes to
the subleading term: 
\begin{equation}
\negthickspace\negthickspace\sigma_{\mathrm{HL}}=1+\frac{X^{+2}}{2h\pi(X^{+2}-1)}\left[\frac{-2}{X^{+2}-1}+\frac{(X^{-2}-1)}{(X^{+}-X^{-})(X^{+}X^{-}-1)}\log\frac{(X^{+}+1)(X^{-}-1)}{(X^{+}-1)(X^{-}+1)}\right]+\mathcal{O}\Big(\frac{1}{h^{2}}\Big).\label{eq:sigmaHL}
\end{equation}

\item The remainder of the dressing phase, $\sigma_{n\geq2}$, does not contribute at this
order; see below for comments. 
\end{itemize}
The final result from the pole at $y^{-}=X^{+}$ is:
\begin{align}
 & \delta E_{(y^{-}=X^{+})}=\delta E_{\mathrm{class.}(y^{-}=X^{+})}+\frac{i\, e^{\frac{i\, p}{2}}e^{-\frac{i\Delta}{h}\frac{X^{+}}{X^{+2}-1}}(X^{+}-X^{-})^{2}}{2(1+X^{-}X^{+})(X^{+2}-1)^{3}X^{-}}\Bigg\{4i\, e^{-\frac{i\, p}{2}}X^{+}(X^{+}X^{-}-1)(X^{+2}-1)\nonumber \\
 & -i\,\left(4X^{+}+X^{+3}+X^{-2}X^{+}(1+4X^{+2})+2X^{-}(1-7X^{+2}+X^{+4})\right)\nonumber \\
 & +\frac{X^{+}}{\pi}\left(2(X^{+}-X^{-})(X^{-}X^{+}-1)+(X^{-2}-1)(X^{+2}-1)\log\left[\frac{(X^{-}-1)(1+X^{+})}{(1+X^{-})(X^{+}-1)}\right]\right)\nonumber \\
 & +\frac{\Delta}{h}\frac{X^{+}(X^{+}-X^{-})(X^{+}X^{-}-1)(1+X^{+2})}{(X^{+2}-1)}\Bigg\}\label{eq:subleading-mu-one-pole}
\end{align}

The pole at $y^{+}=X^{+}$ gives a similar contribution. Adding the two, and taking
into account the sign as before, 
\begin{align}
 & \delta E_{(y^{-}=X^{+})}-\delta E_{(y^{+}=X^{+})}=\delta E_{\mathrm{class.}}+\nonumber \\
 & \quad+\frac{e^{-\frac{i\Delta}{h}\frac{X^{+}}{X^{+2}-1}}(X^{+}-X^{-})^{2}}{X^{-}(X^{+2}-1)^{2}X^{+}(X^{+}X^{-}+1)}\Bigg\{-2X^{+}(X^{-}X^{+}-1)+e^{\frac{i\, p}{2}}X^{-}(X^{+2}-1)\nonumber \\
 & +\frac{i\, e^{\frac{i\, p}{2}}}{\pi}\frac{X^{+}}{X^{-}(X^{+2}-1)}\left(2(X^{+}-X^{-})(X^{-}X^{+}-1)+\left(X^{-2}-1\right)\left(X^{+2}-1\right)\log\frac{(X^{-}-1)(1+X^{+})}{(1+X^{-})(X^{+}-1)}\right)\Bigg\}\label{eq:subleading-mu-final}
\end{align}
which is the leading and subleading $\mu$-term corrections to the energy of the
dyonic elementary magnon.

Let us make a two comments about the contribution of the dressing phase:
\begin{itemize}
\item Our calculation follows that of \cite{Hatsuda:2008gd} in keeping only the AFS and
HL phases, i.e. the terms $\chi$$^{(n)}(x,y)/h^{n-1}$ for $n=0,1$ in \eqref{eq:sigma-exp-chi}.
But if one considers the full sum infinite sum, the terms apparently subleading in
$1/h$ can have non-trivial pole contributions, and thus contribute along with the
earlier terms. Nevertheless, in the present dyonic case, if we perform a careful
calculation of all the terms (following \cite{Janik:2007wt}), we find that the contribution
from all higher terms does indeed vanish. 
\item However if we were to consider a non-dyonic magnon, i.e. to take $Q=1$ for the physical
particle, then there would be contributions from $\sigma_{n\ge2}$ both at leading
and subleading orders. (This is also what happens in the calculation of \cite{Janik:2007wt}.)
Looking at $\sigma_{\mathrm{AFS}}$ \eqref{eq:sigmaAFS}, and likewise \eqref{eq:str},
\eqref{eq:sigmaHL}, it seems that the non-dyonic limit of these expressions is ill
defined, but in fact the divergences should cancel with extra terms that come from
the taking the principal value in the $F$-term. See appendix \ref{sec:S5} for a
discussion of this subject in the $AdS_{5}\times S^{5}$ case.%
\footnote{As noted on page \pageref{paragraph-non-d-mu-puzzles} above, we leave the calculation
of the $\mu$-term for a non-dyonic magnon in $AdS_{4}\times CP^{3}$ for future
work. %
}
\end{itemize}

\section{F-terms for Composite Magnons\label{sec:F-terms-for-Composite-Magnons}}

So far we have studied only the elementary giant magnon, and its dyonic generalisation.
There are various other giant magnon solutions possible in $CP^{3}$, all of which
are superpositions of two elementary magnons. From the sigma-model point of view
this was shown by the dressing construction of \cite{Hollowood:2009sc}. They are
of two kinds:
\begin{itemize}
\item The magnons in $RP^{2}$ and $RP^{3}$ are the same solutions as exist in $S^{2}$
\cite{Hofman:2006xt} and $S^{3}$ \cite{Dorey:2006dq}, and the same is true of
the corresponding finite-$J$ solutions \cite{Abbott:2008qd}. This equivalence also
holds to some degree in both algebraic curve and L\"{u}scher calculations, where
many formulae reduce to exactly what was used in $AdS_{5}\times S^{5}$. \smallskip \smallskip \\
From the S-matrix point of view, the $RP^{2}$ magnon is a superposition of one A-particle
and one B-particle, both with the same $su(2\vert2)$ label which we take to be $a=1$.
In the dyonic case, these are each replaced by bound states of $Q$ particles. 
\item The so-called `big giant magnon' is another superposition of two dyonic magnons,
oriented such that their charges $Q$ cancel out. This was first known in the algebraic
curve \cite{Shenderovich:2008bs} and later constructed using dressing by \cite{Hollowood:2009tw,Kalousios:2009mp,Suzuki:2009sc}.
The most explicit construction of it from two elementary magnons was given by \cite{Hatsuda:2009pc},
from which it is clear that the two particles carry different labels: say A with
$a=1$, plus B with $a=2$. \smallskip \smallskip \\
Classically, when $Q\to1$ this becomes indistinguishable from the $RP^{2}$ solution
(in both sigma-model and algebraic curve). We see this behaviour also for F-terms
we calculate here. Even though they are corrections at order $h^{0}$ (while the
classical energy is order $h$), they are the leading terms in $1/h$ in the coefficient
$a_{m,0}$ of the exponential, \eqref{eq:intro-F1-F2} or \eqref{eq:a_mn-expansion}. 
\end{itemize}
We summarise all of the bound-state and composite magnons in table \ref{tab:bound-and-composite}.

\begin{table}
$\negthickspace\negthickspace$%
\begin{tabular}{cccc|cc}
\multicolumn{4}{c|}{$\vphantom{\dfrac{1}{1}}$Physical $\qquad$} & \multicolumn{2}{c}{$\negthickspace\negthickspace$Mirror }\tabularnewline
$\negthickspace$Elementary$\negthickspace\negthickspace$ & Dyonic & $\negthickspace$Composite $RP^{3}$ & Composite `big' & Light & Heavy\tabularnewline
\hline 
$\vphantom{1^{1^{1}}}A_{a=1}$ & $(A_{a=1})^{Q}$  & $(A_{a=1}B_{a=1})^{Q}$ & $(A_{a=1}B_{a=2})^{Q}$ & $A_{a}$ and $B_{a}$ & $(A_{a}B_{b})$\tabularnewline
 &  &  &  & $a=1,2,3,4.$ & $a,b$ as shown in \eqref{eq:heavy-list}.$\negthickspace$\tabularnewline
\end{tabular}

\caption{Summary of the various states considered, in spin-chain language. It would clearly
be equivalent to consider corrections to, for instance, the $A_{a=2}$ physical particle
or the $(B_{a=2})^{Q}$ dyonic magnon. \label{tab:bound-and-composite}}
\end{table}

In \cite{Abbott:2010yb} we did not write all of the the F-terms which we calculate
here in full. But we give in appendix \ref{sec:AC-Formulae} enough formulae to check
them.

\subsection{The $RP^{3}$ magnon\label{sub:RP3}}

Note that both constituent particles will always be described by the same $X^{\pm}$.
We will use $p$, $E$, $Q$ and $J$ to be the quantities defined in terms of this
in appendix \ref{sub:Parameters-and-E-Q-J-defn}, even though the total momentum
is then $p_{\mathrm{tot}}=2p$, total energy $2E$, and the total charge $2Q$ (or,
for the big magnon, zero). This implies in particular that the kinetic factor will
be unchanged: we have $E_{\mathrm{tot}}'(p_{\mathrm{tot}})=\mathcal{E}'_{Q}(p)$. 

Allowing that the virtual particle may be A or B, the remaining factor of the integrand
in \eqref{eq:dE-F-basic} is the following: 
\begin{align*}
e^{-iq_{\star}L}\sum_{b=1}^{4}(-1)^{F_{b}}\left[S_{b}\tilde{S}_{b}+\tilde{S}_{b}S_{b}\right](y^{\pm},x^{\pm}) & =2\left[(a_{1})^{2}+(a_{1}+a_{2})^{2}-2(a_{6})^{2}\right]n\tilde{n}\sigma^{2}(y^{\pm},x^{\pm})\\
 & =4\, e^{-\frac{i}{h}\frac{x}{x^{2}-1}(2E-1-L)}\left[a_{1}(x,x^{\pm})^{2}-1\right]+\mathcal{O}\Big(\frac{1}{h}\Big)\\
 & =F_{\mathrm{light}\frp}^{(\ell=1)}\Big\vert_{Q=1}+\ldots\,.
\end{align*}
For $\sigma$ we use \eqref{eq:simplest-sigma-n} above, and for all the other factors
we are simply evaluating at $y^{\pm}=x$. For the exponents to match, we must remember
that $\alpha=\Delta_{\mathrm{tot}}/h$ in the algebraic curve contains the total
energy, $\Delta_{\mathrm{tot}}=2E-J$. And, following \eqref{eq:L-equals-J/2}, we
set $L$ to be half the total angular momentum: 
\[
L=J\,.
\]

For the dyonic case (i.e. a superposition of two physical bound states, each with
$Q$ particles) clearly we want instead:
\begin{align*}
e^{-iq_{\star}L}\sum_{b=1}^{4}(-1)^{F_{b}}\left[S_{b}\tilde{S}_{b}+\tilde{S}_{b}S_{b}\right](y^{\pm},X^{\pm}) & =2\, S_{\mathrm{BDS}}\left[(s_{1})^{2}+(s_{2})^{2}-(s_{3})^{2}-(s_{4})^{2}\right]H^{2}\sigma^{2}(y^{\pm},X^{\pm})\\
 & =F_{\mathrm{light}\frp}^{(\ell=1)}+\mathcal{O}\Big(\frac{1}{h}\Big).
\end{align*}
Here we write $H(y^{\pm},X^{\pm})=\sqrt{X^{+}/X^{-}}(y^{-}/y^{+})^{Q/2}$ for compactness,
\eqref{eq:defn-H-appendix}.

\subsubsection*{Second F-term}

As for the elementary magnon, there are two terms which contribute to $\delta E^{F,2}$.
One comes from a light mode (of type A or B) circling the space twice, what we called
the HJL term \eqref{eq:dE-HJL-m}. For this, \eqref{eq:sum-S-like-HJL} is replaced
by
\begin{align*}
e^{-2iq_{\star}L}\sum_{b=1}^{4}(-1)^{F_{b}}\left[(S_{b}\tilde{S}_{b})^{2}+(\tilde{S}_{b}S_{b})^{2}\right](y^{\pm},x^{\pm}) & =2\left[(a_{1})^{4}+(a_{1}+a_{2})^{4}-2(a_{6})^{4}\right](n\tilde{n})^{2}\sigma^{4}(y^{\pm},x^{\pm})\\
 & =4\, e^{-2\frac{i}{h}\frac{x}{x^{2}-1}(2E-1-L)}\left[a_{1}(x,x^{\pm})^{4}-1\right]+\mathcal{O}\Big(\frac{1}{h}\Big)\\
 & =F_{\mathrm{light}\frp}^{(\ell=2)}\Big\vert_{Q=1}+\ldots
\end{align*}
or, in the dyonic case, 
\begin{align*}
e^{-2iq_{\star}L}\sum_{b=1}^{4}(-1)^{F_{b}}\left[(S_{b}\tilde{S}_{b})^{2}+(\tilde{S}_{b}S_{b})^{2}\right](y^{\pm},X^{\pm}) & =2\,(S_{\mathrm{BDS}})^{2}\left[(s_{1})^{4}+(s_{2})^{4}-(s_{3})^{4}-(s_{4})^{4}\right]H^{4}\sigma^{4}(y^{\pm},X^{\pm})\\
 & =F_{\mathrm{light}\frp}^{(\ell=2)}+\mathcal{O}\Big(\frac{1}{h}\Big).
\end{align*}

The other term comes from a heavy mode (or $M=2$ mirror bound state) running in
the loop. Using the bound-state S-matrix $S_{2-1}$ we write
\begin{align*}
e^{-2iq_{\star}L}\sum_{b=1}^{8}(-1)^{F_{b}}S_{b}(Y^{\pm},x^{\pm})S_{b}(Y^{\pm},x^{\pm}) & =\left[\vphantom{\frac{1}{1}}3(a_{5}^{5})^{2}+(2a_{8}^{8})^{2}-2(a_{9}^{9})^{2}-2\Big(\frac{a_{9}^{9}+a_{3}^{3}}{2}\Big)^{2}\right]n\tilde{n}\:\sigma^{2}(Y^{\pm},x^{\pm})\\
 & =F_{\mathrm{heavy}\frp}^{(\ell=1)}\Big\vert_{Q=1}+\mathcal{O}\Big(\frac{1}{h}\Big).
\end{align*}
The dyonic version of this needs $S_{2-Q}$ as constructed in appendix \ref{sub:Constructing-S_2Q-by-fusion},
\begin{align*}
 & e^{-2iq_{\star}L}\sum_{b=1}^{8}(-1)^{F_{b}}S_{b}(Y^{\pm},X^{\pm})S_{b}(Y^{\pm},X^{\pm})\\
 & =e^{-2iq_{\star}L}(T_{0})^{2}\left[\vphantom{\frac{1}{1}}3(t_{1})^{2}+(t_{4})^{2}-2(t_{5})^{2}-2(t_{7})^{2}\right]\sigma^{2}(Y^{\pm},X^{\pm})\displaybreak[0]\\
 & =e^{-\frac{i}{h}\frac{x}{x^{2}-1}(2E-Q-L)}\Big[3-2\left(\frac{x-X^{-}}{x-X^{+}}\sqrt{\frac{X^{+}}{X^{-}}}\right)^{2}+\left(\frac{(x-X^{-})(1-xX^{+})}{(x-X^{+})(1-xX^{-})}\right)^{2}-\left(\frac{1-xX^{+}}{1-xX^{-}}\sqrt{\frac{X^{-}}{X^{+}}}\right)^{2}\Big]+\ldots\\
 & =F_{\mathrm{heavy\frp}}^{(\ell=1)}+\mathcal{O}\Big(\frac{1}{h}\Big).
\end{align*}

\subsection{The big giant magnon\label{sub:Big}}

The only difference from the $RP^{3}$ case above is that one of the external particles
is now $a=2$. Thus we will need for the first time matrix elements $S_{b2}^{b2}$.
For the non-dyonic case these can be read off from \eqref{eq:full-S-hat} as follows:
\[
\hat{S}_{b2}^{b2}=\hat{S}_{b'1}^{b'1},\qquad(b,b')=(1,2),\:(2,1),\:(3,3),\:(4,4)\,.
\]
Then we have 
\begin{align*}
e^{-iq_{\star}L}\sum_{b=1}^{4}(-1)^{F_{b}}\left(S_{b1}^{b1}\tilde{S}_{b2}^{b2}+\tilde{S}_{b1}^{b1}S_{b2}^{b2}\right)(y^{\pm},x^{\pm}) & =e^{-iq_{\star}L}2\left[a_{1}(a_{1}+a_{2})+(a_{1}+a_{2})a_{1}-2a_{6}a_{6}\right]n\tilde{n}\,\sigma^{2}\\
 & =F_{\mathrm{light}\fbig}^{(\ell=1)}\Big\vert_{Q=1}+\mathcal{O}\Big(\frac{1}{h}\Big).
\end{align*}
This agrees with the $RP^{2}$ expression above. 

For the dyonic case, we need the corresponding elements of $S_{Q-1}$. The above
permutation clearly commutes with the fusion procedure used to construct this, and
thus we have
\begin{align*}
e^{-iq_{\star}L}\sum_{b=1}^{4}(-1)^{F_{b}}\left(S_{b1}^{b1}\tilde{S}_{b2}^{b2}+\tilde{S}_{b1}^{b1}S_{b2}^{b2}\right)(y^{\pm},X^{\pm}) & =2\, S_{\mathrm{BDS}}\left[s_{1}s_{2}+s_{2}s_{1}-s_{3}s_{3}-s_{4}s_{4}\right]H^{2}\sigma^{2}(y^{\pm},X^{\pm})\\
 & =F_{\mathrm{light}\fbig}^{(\ell=1)}+\mathcal{O}\Big(\frac{1}{h}\Big).
\end{align*}

\subsubsection*{Second F-term}

The term from a light mode wrapping twice is clearly given by 
\begin{align*}
 & e^{-2iq_{\star}L}\sum_{b=1}^{4}(-1)^{F_{b}}\left[\left(S_{b1}^{b1}\tilde{S}_{b2}^{b2}\right)^{2}+\left(\tilde{S}_{b1}^{b1}S_{b2}^{b2}\right)^{2}\right](y^{\pm},x^{\pm})\\
 & =e^{-2iq_{\star}L}2\left[2(a_{1})^{2}(a_{1}+a_{2})^{2}-2(a_{6})^{4}\right](n\tilde{n})^{2}\sigma^{4}(y^{\pm},x^{\pm})\\
 & =F_{\mathrm{light}\fbig}^{(\ell=2)}\Big\vert_{Q=1}+\mathcal{O}\Big(\frac{1}{h}\Big)
\end{align*}
and for the dyonic case, 
\begin{align*}
 & e^{-2iq_{\star}L}\sum_{b=1}^{4}(-1)^{F_{b}}\left[\left(S_{b1}^{b1}\tilde{S}_{b2}^{b2}\right)^{2}+\left(\tilde{S}_{b1}^{b1}S_{b2}^{b2}\right)^{2}\right](y^{\pm},X^{\pm})\\
 & =e^{-2iq_{\star}L}2(S_{\mathrm{BDS}})^{2}\left[(s_{1}s_{2})^{2}+(s_{2}s_{1})^{2}-(s_{3}s_{3})^{2}-(s_{4}s_{4})^{2}\right]H^{4}\sigma^{4}(y^{\pm},X^{\pm})\\
 & =F_{\mathrm{light}\fbig}^{(\ell=2)}+\mathcal{O}\Big(\frac{1}{h}\Big).
\end{align*}

For the heavy mode, we need $S_{2-1}(Y^{\pm},x^{\pm})_{b2}^{b2}$. Here we make the
following conjecture, based on the list of virtual bound states \eqref{eq:heavy-list},
and the S-matrix \eqref{eq:S-ABBN}: under $a=1\leftrightarrow a=2$, the heavy boson
$b=4$ is not mixed with the other three $b=1,2,3$. (These three all have the same
$S_{b1}^{b1}$.) Meanwhile, the heavy fermions $b=5,6$ and $b=7,8$ are swopped.
We can write this as the following permutation: 
\begin{equation}
\hat{S}_{b2}^{b2}=\hat{S}_{b'1}^{b'1},\qquad b'=\rho(b)\mbox{ where }\rho=(5,7)(6,8)\,.\label{eq:S_b2-M2-conjecture}
\end{equation}
Then using this, we are led to 
\begin{align*}
e^{-2iq_{\star}L}\sum_{b=1}^{8}(-1)^{F_{b}}S_{b}(Y^{\pm},x^{\pm}) & =\left[3(a_{5}^{5})^{2}+(2a_{8}^{8})^{2}-4a_{9}^{9}\frac{a_{9}^{9}+a_{3}^{3}}{2}\right]\, n\tilde{n}\:\sigma^{2}\\
 & =F_{\mathrm{heavy}\fbig}^{(\ell=1)}\Big\vert_{Q=1}+\mathcal{O}\Big(\frac{1}{h}\Big).
\end{align*}
This conjecture follows through to the dyonic case, where in terms of $S_{2-Q}$
we have: 
\begin{align}
 & e^{-2iq_{\star}L}\sum_{b=1}^{8}(-1)^{F_{b}}\left(S_{b1}^{b1}\tilde{S}_{b2}^{b2}+\tilde{S}_{b1}^{b1}S_{b2}^{b2}\right)(Y^{\pm},X^{\pm})\nonumber \\
 & =e^{-2iq_{\star}L}(T_{0})^{2}\left[3(t_{1})^{2}+(t_{4})^{2}-4(t_{5}t_{7})\right]\sigma^{2}(Y^{\pm},X^{\pm})\displaybreak[0]\nonumber \\
 & =e^{-\frac{i}{h}\frac{x}{x^{2}-1}(2E-Q-L)}\frac{(x^{2}-1)(X^{-}-X^{+})\left[x^{2}(3X^{-}-X^{+})+3X^{+}-X^{-}-2x(X^{-}X^{+}+1)\right]}{(xX^{-}-1)^{2}(x-X^{+})^{2}}+\ldots\nonumber \\
 & =F_{\mathrm{heavy}\fbig}^{(\ell=1)}+\mathcal{O}\Big(\frac{1}{h}\Big).\label{eq:F-big-AND-heavy-dyonic}
\end{align}

The interpretation of the big giant magnon used here differs from that in \cite{Ahn:2010eg}.
There, it was treated as a superposition of a magnon and an `anti-magnon', both of
type A, with the latter defined by sending $X^{\pm}\to1/X^{\mp}$ (so as to send
$Q\to-Q$). This appears to give to the correct first and second F-terms: writing
a prime for the `anti-magnon', in the non-dyonic case these read: 
\begin{align*}
F_{\mathrm{light}\fbig}^{(\ell=1)}\Big\vert_{Q=1} & =\left[a_{1}a_{1}'+(a_{1}+a_{2})(a_{1}+a_{2}')-2a_{6}a_{6}'\right](nn'+\tilde{n}\tilde{n}')\sigma\sigma'+\mathcal{O}\Big(\frac{1}{h}\Big)\\
F_{\mathrm{light}\fbig}^{(\ell=2)}\Big\vert_{Q=1} & =\left[(a_{1}a_{1}')^{2}+(a_{1}+a_{2})^{2}(a_{1}+a_{2}')^{2}-2(a_{6}a_{6}')^{2}\right]\left[(nn')^{2}+(\tilde{n}\tilde{n}')^{2}\right](\sigma\sigma')^{2}+\ldots\\
F_{\mathrm{heavy}\fbig}^{(\ell=1)}\Big\vert_{Q=1} & =\left[3a_{5}^{5}a_{5}^{5\prime}+4a_{8}^{8}a_{8}^{8\prime}-2a_{9}^{9}a_{9}^{9\prime}-(a_{9}^{9}+a_{3}^{3})(a_{9}^{9\prime}+a_{3}^{3\prime})\right]n\tilde{n}\: n'\tilde{n}'\,\sigma\sigma'+\ldots\,.
\end{align*}
All of these match the $RP^{2}$ magnon. In the dyonic case, we can recover the algebraic
curve results if we use in the S-matrix for the `anti-magnon' the following prefactor:%
\footnote{We thank Minkyoo Kim for discussions of this point.%
}
\begin{align*}
S_{\mathrm{BDS}}'(y^{\pm},X^{\pm}) & =S_{\mathrm{BDS}}(y^{\pm},1/X^{\mp})
\end{align*}
Note however that this is not equal to $\prod_{k=1}^{Q}s_{\mathrm{BDS}}(y^{\pm},1/x_{k}^{\mp})$.
In this calculation we evaluate the S-matrix at $y^{\pm}=y$; a more strict test
of this interpretation could be obtained by performing a calculation in a regime
where $y^{+}\neq y^{-}$, i.e. at the next order in $1/h$.

\subsection{Agreement to all orders}

As we noted for the elementary giant magnons in section \ref{sub:All-higer-F-terms},
these results can be extended to recover all higher F-terms. The total contribution
from elementary virtual particles matches that from the light modes in the algebraic
curve, thanks to the following relations:
\begin{align*}
\prod_{b}\left[1-S_{b}\tilde{S}_{b}e^{-iq_{*}L}\right]^{(-2)^{F_{b}}} & =\prod_{\stackrel{ij}{\mathrm{light[RP]}}}\left[1-e^{-i(q_{i}-q_{j})}\right]^{(-1)^{F_{ij}}}+\mathcal{O}\Big(\frac{1}{h}\Big)\\
\prod_{b}\left[(1-S_{b1}\tilde{S}_{b2}e^{-iq_{*}L})(1-\tilde{S}_{b1}S_{b2}e^{-iq_{*}L})\right]^{(-1)^{F_{b}}} & =\prod_{\stackrel{ij}{\mathrm{light[Big]}}}\left[1-e^{-i(q_{i}-q_{j})}\right]^{(-1)^{F_{ij}}}+\mathcal{O}\Big(\frac{1}{h}\Big).
\end{align*}
The contribution from the virtual bound states (of only $M=2$ particles) similarly
matches that from the heavy modes:
\begin{align*}
\prod_{b}\left[1-S_{b}(Y^{\pm},X^{\pm})e^{-2iq_{*}L}\right]^{(-2)^{F_{b}}} & =\prod_{\stackrel{ij}{\mathrm{heavy[RP]}}}\left[1-e^{-i(q_{i}-q_{j})}\right]^{(-1)^{F_{ij}}}+\mathcal{O}\Big(\frac{1}{h}\Big)\\
\prod_{b}\left[(1-S_{b1}\tilde{S}_{b2})(Y^{\pm},X^{\pm})e^{-2iq_{*}L})(1-\tilde{S}_{b1}S_{b2}(Y^{\pm},X^{\pm})e^{-2iq_{*}L})\right]^{(-1)^{F_{b}}}\hspace{-4cm}\\
 & =\prod_{\stackrel{ij}{\mathrm{heavy[Big]}}}\left[1-e^{-i(q_{i}-q_{j})}\right]^{(-1)^{F_{ij}}}+\mathcal{O}\Big(\frac{1}{h}\Big).
\end{align*}
We have thus recovered the exact F-term (i.e. to all orders $m$) at one loop for
all types of $CP^{3}$ giant magnon.

\section{Conclusions\label{sec:Conclusions}}

The complete energy of the giant magnon, including all exponential corrections, is
given by \cite{Gromov:2008ie} 
\begin{equation}
E=\negthickspace\negthickspace\sum_{m,n=0,1,2\ldots}\negthickspace\negthickspace a_{m,n}\left(e^{-\Delta/2h}\right)^{m}\left(e^{-\Delta/E}\right)^{n}.\label{eq:a_mn-expansion}
\end{equation}
Here each coefficient $a_{m,n}$ is a function of the coupling. Our previous paper
\cite{Abbott:2010yb} calculated the infinite-volume term $a_{0,0}=\mathcal{E}_{Q}(p)$,
as well as F-terms $a_{1,0}$ and $a_{2,0}$, at one loop in the string theory. We
used the algebraic curve formulation, which exploits the integrability of the classical
string theory, but is otherwise just a re-formulation of the sigma-model. 

In this paper we have shown that, for L\"{u}scher corrections, the heavy modes appearing
in the string theory can be identified as bound states in the mirror theory. This
allowed us to correctly recover the second F-term correction $a_{2,0}$ first computed
in \cite{Abbott:2010yb}. (The first F-term $a_{1,0}$ was also calculated by \cite{Shenderovich:2008bs,Bombardelli:2008qd}
and for the dyonic case \cite{Ahn:2010eg}.) When treating dyonic giant magnons,
we needed to scatter a physical bound state (of $Q$ particles) with a virtual bound
state (of two), for which we were able to derive the appropriate S-matrix. 

We have also computed $\mu$-term corrections for the elementary dyonic giant magnon.
The leading term matches the classical finite-$J$ correction for this magnon, first
given by \cite{Abbott:2009um}. We went on to compute the subleading term, giving
a one-loop prediction. Since it comes from the all-loop S-matrix, this prediction
is the second term in an expansion in $1/h(\lambda)$. This is related to the string
theory's coupling constant $\lambda\propto R^{2}/\alpha'$ (at strong coupling) by
\begin{equation}
h(\lambda)=\sqrt{\frac{\lambda}{2}}+c+\mathcal{O}\Big(\frac{1}{\sqrt{\lambda}}\Big).\label{eq:h-strong}
\end{equation}
Because the F-term corrections vanish classically (order $h\sim\sqrt{\lambda}$),
their order $1$ terms are sensitive only to the leading term in the expansion \eqref{eq:h-strong}.
But, like the infinite-volume energy $a_{0,0}$, the one-loop part of the $\mu$-term
is subleading in $1/h$. One extension which we hope to address in a forthcoming
paper is the corresponding string theory calculation of $a_{0,1}$ at one loop, along
the lines of \cite{Gromov:2008ec}. 

Finally, we were also able to extenbd our calculation to obtain all higher F-terms
$a_{m,0}$. In this calculation the elementary virtual particles contribute to every
term via \cite{Heller:2008at}'s generalised L\"{u}scher formula \eqref{eq:dE-HJL-m},
by circling the space $m$ times. The two-particle virtual bound states ($M=2$)
contribute to even-numbered terms $m=2\ell$ in the same way. We did not include
any bound states of $M\geq3$ particles; this is natural from the algebraic curve
perspective where one has modes of mass 1 and 2 only, but does not seem obvious from
the S-matrix (see section \ref{sub:All-higer-F-terms}).

In order to understand this better, a possible connection of these L\"{u}scher corrections
with the asymptotic solution of the $AdS_{4}/CFT_{3}$ TBA/Y-system proposed in \cite{Bombardelli:2009xz,Gromov:2009at}
is currently under investigation. It would be very interesting to understand from
this how the F-terms involving all bound states at weak coupling re-sum to involve
only $M=1,2$ at strong coupling.

\subsection*{Acknowledgements}

We have benefitted greatly from conversations with Changrim Ahn, Zoltan Bajnok, Davide
Fioravanti, Minkyoo Kim, Rafael Nepomechie, and especially with Olof Ohlsson Sax,
and it is a pleasure to thank them all. We would also like to thank a J. Phys. A
referee for some very useful comments. 

For hospitality while working on this, we thank the Perimeter Institute, Nordita,
and Humboldt University Berlin. I. Aniceto was partly funded by Funda\c{c}\~ao para
a Ci\^encia e Tecnologia, fellowship SFRH\slash{}BPD\slash{}69696\slash{}2010.
D. Bombardelli was partly funded by Funda\c{c}\~ao para a Ci\^encia e Tecnologia,
fellowship SFRH\slash{}BPD\slash{}69813\slash{}2010, and also by the network UNIFY
for travel support.

\appendix

\section{$\mu$-terms for the $AdS_{5}\times S^{5}$ Giant Magnon\label{sec:S5}}

We show here the analogue in $AdS_{5}\times S^{5}$ of our calculation of the $\mu$-terms
for a dyonic giant magnon in $AdS_{4}\times CP^{3}$ in section \ref{sub:Subleading-mu-term}. 

The dispersion relation for this theory is as follows: 
\begin{align*}
\Delta-J=\mathcal{E}^{AdS_{5}}(p) & =\sqrt{Q^{2}+\frac{\lambda}{\pi^{2}}\sin^{2}\frac{p}{2}}\\
 & =2\mathcal{E}_{Q}(p)\qquad\qquad\qquad\qquad\mbox{with }h=\smash{\frac{\sqrt{\lambda}}{4\pi}}\\
 & =-ih\left(X^{+}-\frac{1}{X^{+}}-X^{-}+\frac{1}{X^{-}}\right).
\end{align*}
With this identification of $h$ and of $\mathcal{E}^{AdS_{5}}$ we can now re-use
all of our expressions from section \ref{sub:Setup-for-mu}, multiplying $q$ and
thus the Jacobian factor \eqref{eq:setup-mu-jacobian} by 2. (But leaving the kinetic
term \eqref{eq:setup-mu-kinetic} unchanged.) 

The important difference is the S-matrix involved, from which instead of \eqref{eq:sum-S-dyonic-mu}
we should use:
\begin{align}
\sum_{b}S_{b}(y^{\pm},X^{\pm}) & =S_{\mathrm{BDS}}\left[s_{1}+s_{2}-s_{3}-s_{4}\right]^{2}\:\frac{X^{+}}{X^{-}}\left(\frac{y^{-}}{y^{+}}\right)^{Q}\sigma^{2}\,.\label{eq:S5-S-matrix}
\end{align}
All the pieces of this are the same as we used above. Adding the contributions at
both poles $y^{\pm}=X^{+}$, we get the following leading correction: 
\[
\delta E_{\mathrm{class.}}=\frac{2ih}{X^{+}X^{-}}\frac{(X^{+}-X^{-})^{3}}{(X^{+}X^{-}+1)}e^{\frac{X^{+}}{X^{+2}-1}\left(\frac{i}{h}L+\frac{i}{h}Q+\frac{2}{X^{+}}-\frac{2}{X^{-}}\right)}.
\]
This is exactly the term given by \cite{Hatsuda:2008gd}. 

Going to the next order, the contribution from the pole $y^{-}=X^{+}$ is this: 
\begin{align*}
\delta E_{(y^{-}=X^{+})} & =\delta E_{\mathrm{class.}(y^{-}=X^{+})}+\frac{4(X^{+}-X^{-})^{2}e^{ip-i\frac{\Delta X^{+}}{h(X^{+2}-1)}}}{(X^{+}X^{-}+1)((X^{+})^{2}-1)^{2}}\left[-2e^{-i\frac{p}{2}}(X^{+2}-1)+i\frac{X^{+}-X^{-}}{\pi}\right.\\
 & \qquad\frac{X^{+}+X^{-}(1+X^{+2}(-4+X^{+}(X^{-}+X^{+})))}{X^{+}(X^{+}X^{-}-1)}+i\frac{\Delta}{4h}\frac{(X^{+2}+1)(X^{+}-X^{-})}{X^{+}-1}\\
 & \qquad\left.+\frac{i((X^{-})^{2-1})(X^{+2}-1)}{2\pi(X^{+}X^{-}-1)}\log\frac{(X^{-}-1)(X^{+}+1)}{(X^{-}+1)(X^{+}-1)}\right].
\end{align*}
The contribution from the pole at $y^{+}=X^{+}$ is similar, and adding them (with
the same minus as before) 
\begin{align}
\negthickspace\negthickspace & \negthickspace\negthickspace\delta E_{(y^{-}=X^{+})}-\delta E_{(y^{+}=X^{+})}=\delta E_{\mathrm{class.}}+\mbox{Re}\Bigg\{\frac{4(X^{+}-X^{-})^{2}e^{ip-i\frac{\Delta X^{+}}{h((X^{+})^{2}-1)}}}{(X^{+}X^{-}+1)(X^{+2}-1)^{2}}\left[\frac{X^{-}(X^{+2}-1)^{2}}{X^{+}(X^{+}X^{-}-1)}\right.\nonumber \\
\negthickspace\negthickspace & \left.-2e^{-i\frac{p}{2}}((X^{+})^{2}-1)+2i\frac{X^{+}-X^{-}}{\pi}+\frac{i((X^{-})^{2-1})((X^{+})^{2}-1)}{\pi(X^{+}X^{-}-1)}\log\frac{(X^{-}-1)(X^{+}+1)}{(X^{-}+1)(X^{+}-1)}\right]\Bigg\}.\label{eq:S5-subleading}
\end{align}

Naively the non-dyonic limit of this diverges, giving: 
\[
\delta E^{\mu}=-16e^{-\frac{J}{2h\sin\frac{p}{2}}-2}\left[h\sin^{3}\frac{p}{2}+\frac{h\sin\frac{p}{2}}{Q}+\frac{\sin\frac{p}{2}}{\pi}\right].
\]
However there is another contribution from the fact that a pole in the F-term integral
at $x=X^{+}$ approaches the integration contour:
\begin{align*}
\delta E^{F} & =\fint_{\mathbb{U}_{+}}dx\frac{2i\: x}{\pi(x^{2}-1)^{2}}\left(1-\frac{\mathcal{E}'(p)}{\varepsilon'(q_{*})}\right)e^{-\frac{i\Delta x}{h(x^{2}-1)}}\left(e^{i\frac{p}{2}}\frac{x-X^{-}}{x-X^{+}}+e^{i\frac{p}{2}}\frac{x-1/X^{+}}{x-1/X^{-}}-2\right)^{2}\\
 & \to\fint_{\mathbb{U}_{+}}dx\frac{2i\: x}{\pi(x^{2}-1)^{2}}\left(1-\frac{\mathcal{E}'(p)}{\varepsilon'(q_{*})}\right)e^{-\frac{i\Delta x}{h(x^{2}-1)}}\left(2e^{i\frac{p}{2}}\frac{x-1/x^{+}}{x-x^{+}}-2\right)^{2}\\
 & \qquad+e^{-\frac{J}{2h\sin\frac{p}{2}}-2}\left(-\frac{16h\sin^{3}\frac{p}{2}}{Q}+\frac{4iJ\cos\frac{p}{2}}{h}-8i\sin\frac{p}{2}+8i\sin p\right)+\mathcal{O}\Big(\frac{1}{h}\Big).
\end{align*}
This clearly cancels the $1/Q$ term, and taking the real part we get the following
total non-dyonic correction: 
\begin{equation}
\delta E=-16\, e^{-\frac{J}{2h\sin\frac{p}{2}}-2}\left[h\sin^{3}\frac{p}{2}+\frac{\sin\frac{p}{2}}{\pi}+\mathcal{O}\Big(\frac{1}{h}\Big)\right].\label{eq:S5-subleading-non-d}
\end{equation}
This agrees with the L\"{u}scher correction calculated by \cite{Gromov:2008ec}
by considering the non-dyonic case from the beginning. This is a nontrivial agreement
as in that case there are contributions from higher terms in the dressing phase $\sigma_{n\geq2}$,
which did not enter into our derivation of the the dyonic case above.

This non-dyonic subleading term \eqref{eq:S5-subleading-non-d} also matches the
real part of the algebraic curve result in \cite{Gromov:2008ec}. (The real part
would be obtained there if one included also the contributions in the lower half-plane.)
However the dyonic subleading term \eqref{eq:S5-subleading} differs from the algebraic
curve result of \cite{Gromov:2008ec}.

\section{Two-particle and bound-state S-matrices\label{sec:S-matrices}}

In this appendix we collect familiar formulae for magnons and their S-matrix. We
discuss a number of kinds of bound-state S-matrices: $S_{1-Q}$ ($Q\sim h$ physical),
$S_{M-1}$ ($M\sim1$ virtual), and $S_{M-Q}$ (mixed).

\subsection{Parameters\label{sub:Parameters-and-E-Q-J-defn}}

We describe magnons using Zhukovsky variables (in the complex spectral plane) which
are defined in terms of the charge $Q$ and momentum $p$ by 
\begin{align}
Q(X^{\pm}) & =-ih\left(X^{+}+\frac{1}{X^{+}}-X^{-}-\frac{1}{X^{-}}\right)\label{eq:defn-Xp-Xm-Qp}\\
p & =-i\:\log\frac{X^{+}}{X^{-}}\nonumber 
\end{align}
or, solving for $X^{\pm}$, 
\[
X^{\pm}=e^{\pm ip/2}\frac{\frac{Q}{2}+\sqrt{\frac{Q^{2}}{4}+4\, h^{2}\:\sin^{2}\frac{p}{2}}}{2h\:\sin\frac{p}{2}}\,.
\]
The dispersion relation can be written in terms of these:
\begin{align}
E & =\mathcal{E}_{Q}(p)=\sqrt{\frac{Q^{2}}{4}+4\, h^{2}\:\sin^{2}\frac{p}{2}}\label{eq:disp-rel-Q}\\
 & =-i\frac{h}{2}\left(X^{+}-\frac{1}{X^{+}}-X^{-}+\frac{1}{X^{-}}\right).\nonumber 
\end{align}
We will need the derivative of this with respect to $p$, holding $Q$ fixed, which
can be written as
\begin{equation}
\mathcal{E}_{Q}'(p)=h\frac{X^{+}+X^{-}}{X^{+}X^{-}+1}\,.\label{eq:Eprime-in-XpXm}
\end{equation}

While we have given these initial formulae for the general dyonic case, we will discuss
first the case of just one magnon, $Q=1$, for which we use lower-case $x^{\pm}$
and $\varepsilon(p)$. Then we can expand in $1/h$ to write:
\begin{align*}
\varepsilon(p) & =2h\,\sin\frac{p}{2}+\mathcal{O}\Big(\frac{1}{h}\Big)\\
\varepsilon'(p) & =h\,\cos\frac{p}{2}+\mathcal{O}\Big(\frac{1}{h}\Big)\\
x^{\pm} & =e^{\pm ip/2}\left[1+\frac{1}{4h\,\sin\frac{p}{2}}+\mathcal{O}\Big(\frac{1}{h^{2}}\Big)\right].
\end{align*}

\subsection{Two-particle S-matrix\label{sub:Two-particle-S-matrix}}

Now consider two particles $x^{\pm}$ and $y^{\pm}$. The ABJM S-matrix is \cite{Ahn:2008aa}
\begin{align*}
S(y^{\pm},x^{\pm}) & \;=S_{AA}=S_{BB}\;=\hat{S}\: n\,\sigma\\
\tilde{S}(y^{\pm},x^{\pm}) & \;=S_{AB}=S_{BA}\;=\hat{S}\:\tilde{n}\,\sigma\,.
\end{align*}
with the same $su(2|2)$ invariant matrix part $\hat{S}$ \cite{Beisert:2005tm,Arutyunov:2006yd}
as the SYM case, but one less power of the BES dressing phase $\sigma$ \cite{Beisert:2006ez}.
Explicitly, in our notation the SYM case has: 
\[
S=\hat{S}\otimes\hat{S}\: n\tilde{n}\,\sigma^{2}\,.
\]
The factors which distinguish particles of type A and B \cite{Ahn:2008aa} we have
named $n$ and $\tilde{n}$: 
\[
n(y^{\pm},x^{\pm})=\frac{1-\frac{1}{y^{+}x^{-}}}{1-\frac{1}{y^{-}x^{+}}},\qquad\tilde{n}(y^{\pm},x^{\pm})=\frac{y^{-}-x^{+}}{y^{+}-x^{-}}\,.
\]
Note of course that $\tilde{n}(y^{\pm},x^{\pm})=1/\tilde{n}(x^{\pm},y^{\pm})=1/\tilde{n}(y^{\mp},x^{\mp})$
and similarly for $n(y^{\pm},x^{\pm})$. 

The relevant terms of the matrix part are given by \cite{Arutyunov:2006yd} 
\begin{align}
\hat{S} & =a_{1}\left(E_{11}^{11}+E_{22}^{22}+E_{12}^{12}+E_{21}^{21}\right)+a_{2}\left(E_{12}^{12}+E_{21}^{21}\right)\nonumber \\
 & +a_{3}\left(E_{33}^{33}+E_{44}^{44}+E_{34}^{34}+E_{43}^{43}\right)+a_{4}\left(E_{34}^{34}+E_{43}^{43}\right)\nonumber \\
 & +a_{5}\left(E_{13}^{13}+E_{14}^{14}+E_{23}^{23}+E_{24}^{24}\right)\nonumber \\
 & +a_{6}\left(E_{31}^{31}+E_{41}^{41}+E_{32}^{32}+E_{42}^{42}\right)\qquad+\mbox{ terms }E_{ab}^{ba}\,.\label{eq:full-S-hat}
\end{align}
Particles $b=1,2$ are bosons, and $b=3,4$ here are fermions.%
\footnote{Remembering that each can be type A or B, we have 4+4 particles in all --- exactly
the number of light modes. In the SYM case instead we have $4^{2}=16=8+8$ particles,
i.e. all the transverse modes of the string.%
} The coefficients are:%
\footnote{These are taken from \cite{Arutyunov:2006yd}. In \cite{Janik:2007wt} and some other
papers, $a_{2}$ was given with a factor $(x^{-}-y^{+})$, rather than $+$, which
is an important distinction when it comes to AB bound states. This was the error
corrected in v5 of \cite{Hatsuda:2008gd}, their equation (3.6).%
}
\begin{align}
a_{1}(y^{\pm},x^{\pm}) & =\frac{y^{+}-x^{-}}{y^{-}-x^{+}}\;\sqrt{\frac{x^{+}}{x^{-}}}\:\sqrt{\frac{y^{-}}{y^{+}}} & a_{3}(y^{\pm},x^{\pm}) & =-1\nonumber \\
a_{2}(y^{\pm},x^{\pm}) & =\frac{(y^{-}-y^{+})(x^{-}-x^{+})(x^{-}+y^{+})}{(y^{-}-x^{+})(x^{-}y^{-}-x^{+}y^{+})}\;\sqrt{\frac{x^{+}}{x^{-}}}\:\sqrt{\frac{y^{-}}{y^{+}}} & a_{4}(y^{\pm},x^{\pm}) & =a_{2}(x^{\mp},y^{\mp})\displaybreak[0]\label{eq:a1a2a6}\\
a_{6}(y^{\pm},x^{\pm}) & =\frac{y^{+}-x^{+}}{y^{-}-x^{+}}\;\sqrt{\frac{y^{-}}{y^{+}}} & a_{5}(y^{\pm},x^{\pm}) & =a_{6}(x^{\mp},y^{\mp})\,.\nonumber 
\end{align}
We have assumed (as we will always do) that we are in the string frame. This has
produced the square-root factors, which are the phases $\eta$, often written 
\[
\frac{\eta_{1}}{\tilde{\eta}_{1}}=\sqrt{\frac{x^{+}}{x^{-}}}=e^{ip/2},\qquad\frac{\eta_{2}}{\tilde{\eta}_{2}}=\sqrt{\frac{y^{-}}{y^{+}}}=e^{-i\, q_{\star}/2}.
\]
The alternative is the spin chain frame, which sets both of these to 1. We observe
that in this case, we have exactly $a_{1}=1/\tilde{n}$. 

The BES dressing phase is
\begin{equation}
\sigma(y^{\pm},x^{\pm})=\exp\left[i\chi(y^{-},x^{-})-i\chi(y^{-},x^{+})-i\chi(y^{+},x^{-})+i\chi(y^{+},x^{+})\vphantom{1_{1_{1}}^{1^{1}}}\right].\label{eq:sigma-exp-chi}
\end{equation}
Expanding $\chi=\sum_{n=0}^{\infty}\chi^{(n)}/h^{n-1}$, the leading term gives the
AFS phase, and takes the simple form 
\[
\chi^{(0)}(y,x)=h\left(\frac{1}{y}-\frac{1}{x}\right)\left[1-(1-xy)\log\left(1-\frac{1}{xy}\right)\right].
\]
The next term $\chi^{(1)}$ gives the HL phase. Writing $\chi(x,y)=\tilde{\chi}(x,y)-\tilde{\chi}(y,x)$,
this is given by 
\begin{align*}
\tilde{\chi}^{(1)}(x,y) & =\frac{1}{2\pi}\mathrm{Li}_{2}\left(1-\frac{x+1}{x-1}\frac{y-1}{y+1}\right)-\frac{1}{2\pi}\mathrm{Li}_{2}\left(1-\frac{x+1}{x-1}\frac{y-1}{y+1}e^{i\pi}\right)\\
 & \quad-\frac{i}{2}\mathrm{Log}\left(1-\frac{x+1}{x-1}\frac{y-1}{y+1}e^{i\pi}\right)+\frac{\pi}{8}+\frac{1}{2\pi}\left[\mathrm{Li}_{2}(y)-\mathrm{Li}_{2}(-y)-\mathrm{Li}_{2}(x)+\mathrm{Li}_{2}(-x)\right].
\end{align*}
The expressions for subsequent terms $\chi^{(n)}$ $n\geq2$ can be found in \cite{Janik:2007wt}.

\subsection{Physical bound states: $S_{1-Q}$\label{sub:Physical-bound-states-S}}

If two bosonic particles are of the same type, then in $S^{AA}$ we have a factor
$a_{1}$, which has a pole at $y^{-}=x^{+}$. This pole is what is used to build
the bound states corresponding to dyonic giant magnons. The spectral parameters of
the constituent particles are then connected up like this: 
\begin{equation}
x_{k}^{\pm},\quad k=1,2\ldots Q\qquad\mbox{ with }\; x_{k}^{-}=x_{k-1}^{+}\,.\label{eq:bound-state-overlap-condition}
\end{equation}
From \eqref{eq:disp-rel-Q} it is easy to see that in the total energy, all intermediate
$x_{k}^{\pm}$ cancel out, leaving only 
\[
X^{+}=x_{Q}^{+},\qquad X^{-}=x_{1}^{-}\,.
\]
The total charge $Q$ and momentum $p$ are likewise given in terms of these capital
spectral parameters by \eqref{eq:defn-Xp-Xm-Qp}. 

The S-matrix for scattering of one particle `$b$' off of this bound state is simply
the product of the S-matrices with each constituent particle:
\[
S_{1-Q}(y,X)_{b1}^{b1}=\prod_{k=1}^{Q}S(y,x_{k})_{b1}^{b1}\,.
\]
Fortunately some similar cancellations happen here, again removing dependence on
the individual $x_{k}$. Consider first the case of AA scattering, for which 
\[
S^{AA}(y^{\pm},x^{\pm})_{11}^{11}=\frac{n}{\tilde{n}}\:\sqrt{\frac{x^{+}}{x^{-}}}\sqrt{\frac{y^{-}}{y^{+}}}\:\sigma=s_{0}\,\sigma\,.
\]
The factor $s_{\mathrm{BDS}}=n/\tilde{n}$ also appears in the SYM case in $S_{(11)(11)}^{(11)(11)}=(a_{1})^{2}n\tilde{n}\,\sigma^{2}$.
It has the following cancellation property:
\begin{align}
\prod_{k=1}^{Q}s_{\mathrm{BDS}}(y^{\pm},x_{k}^{\pm}) & =\prod_{k=1}^{Q}\frac{n(y^{\pm},x_{k}^{\pm})}{\tilde{n}(y^{\pm},x_{k}^{\pm})}\nonumber \\
 & =\frac{(y^{+}-X^{-})(1-\frac{1}{y^{+}X^{-}})}{(y^{-}-X^{+})(1-\frac{1}{y^{-}X^{+}})}\frac{(y^{-}-X^{-})(1-\frac{1}{y^{-}X^{-}})}{(y^{+}-X^{+})(1-\frac{1}{y^{+}X^{+}})}\equiv S_{\mathrm{BDS}}(y^{\pm},X^{\pm})\,.\label{eq:S-BDS}
\end{align}
Here we must use $Q=1$ for $x_{k}^{\pm}$ and $y^{\pm}$.%
\footnote{The useful formula is 
\[
(y^{+}-x^{+})\Big(1-\frac{1}{y^{+}x^{+}}\Big)=(y^{-}-x^{-})\Big(1-\frac{1}{y^{-}x^{-}}\Big).
\]
Also using $Q=1$, we have $S_{\mathrm{BDS}}(y^{\pm},x^{\pm})=s_{\mathrm{BDS}}(y^{\pm},x^{\pm})$.%
} More obviously, we have 
\begin{equation}
\prod_{k=1}^{Q}\sqrt{\frac{x_{k}^{+}}{x_{k}^{-}}}\sqrt{\frac{y^{-}}{y^{+}}}=\sqrt{\frac{X^{+}}{X^{-}}}\left(\frac{y^{-}}{y^{+}}\right)^{Q/2}\equiv H(y^{\pm},X^{\pm})\,.\label{eq:defn-H-appendix}
\end{equation}
and for the dressing factor, the antisymmetry $\chi(y,x)=-\chi(x,y)$ leads to another
cancellation: 
\[
\prod_{k=1}^{Q}\sigma(y^{\pm},x_{k}^{\pm})=\sigma(y^{\pm},X^{\pm})\,.
\]

We must still consider matrix elements other than $S^{AA}(y^{\pm},x^{\pm})_{11}^{11}$.
If we define
\[
s_{b}\equiv\prod_{k=1}^{Q}\frac{\hat{S}_{b1}^{b1}}{\hat{S}_{11}^{11}}\,.
\]
then obviously $s_{1}=1$, and some similar cancellations lead to \cite{Hatsuda:2008gd}
\begin{align*}
s_{2} & =\frac{(y^{+}-X^{+})(1-\frac{1}{y^{-}X^{+}})}{(y^{+}-X^{-})(1-\frac{1}{y^{-}X^{-}})}\\
s_{3}=s_{4} & =\frac{y^{+}-X^{+}}{y^{+}-X^{-}}\sqrt{\frac{X^{-}}{X^{+}}}.
\end{align*}
Finally, considering also AB scattering, it is clear that to change the prefactor
from $n$ to $\tilde{n}$ you divide by $s_{\mathrm{BDS}}$, removing that factor.
All together, the final S-matrix is given by
\begin{align*}
S_{1-Q}^{AA}(y^{\pm},X^{\pm})_{b1}^{b1}=S_{1-Q}^{BB}(y^{\pm},X^{\pm})_{b1}^{b1} & =s_{b}\: S_{\mathrm{BDS}}\: H\:\sigma\\
S_{1-Q}^{AB}(y^{\pm},X^{\pm})_{b1}^{b1}=S_{1-Q}^{BA}(y^{\pm},X^{\pm})_{b1}^{b1} & =s_{b}\: H\:\sigma\,.
\end{align*}
with $b=1,2,3,4$.

From this S-matrix, we can get some elements of $S_{Q-1}$ by the following symmetry:

\begin{equation}
S_{Q-1}(Y^{\pm},x^{\pm})_{ba}^{ba}=S_{1-Q}(x^{\mp},Y^{\mp})_{ab}^{ab}\,.\label{eq:reversing-equation}
\end{equation}
Note however that we will only have $S_{Q-1}(Y^{\pm},x^{\pm})_{1a}^{1a}$, since
we started with one particular bound state of $Q$ particles. And further that this
is a physical bound state. Some $S_{Q-Q'}(Y^{\pm},X^{\pm})$ bound-state S-matrices
were constructed by \cite{Hatsuda:2009pc}, for the scattering of two physical dyonic
giant magnons. However for scattering involving bound states of \emph{virtual} particles,
we need something different.

\subsection{Mirror bound states: $S_{M-1}$\label{sub:Mirror-bound-state-S}\label{sub:Bound-states-in-ABBN}}

For a bound state of $M$ particles, \cite{Ahn:2010yv} give (twisted generalisations
of) these S-matrix elements:
\begin{equation}
(-1)^{F_{b}}\hat{S}_{M-1}(Y^{\pm},x^{\pm})_{b1}^{b1}=\begin{cases}
a_{5}^{5}(x^{\pm},Y^{\pm}),\qquad & b=1,\ldots,M+1\\
2a_{8}^{8} & b=M+2,\ldots,2M\\
(-1)\: a_{9}^{9} & b=2M+1,\ldots,3M\\
(-1)\:\dfrac{a_{9}^{9}+a_{3}^{3}}{2} & b=3M+1,\ldots,4M\,.
\end{cases}\label{eq:S-ABBN}
\end{equation}
The coefficients are from \cite{Bajnok:2008qj},%
\footnote{We have corrected the overall sign of $a_{3}^{3}$. The reason these are defined
with arguments $(x^{\pm},Y^{\pm})$ is that appendix $B$ of \cite{Bajnok:2008qj}
is for $S_{1-Q}$. %
} which we write at once in the string frame: 
\begin{align*}
a_{5}^{5}(x^{\pm},Y^{\pm}) & =\frac{x^{+}-Y^{+}}{x^{+}-Y^{-}}\sqrt{\frac{Y^{-}}{Y^{+}}}\\
a_{8}^{8}(x^{\pm},Y^{\pm}) & =\frac{x^{-}(x^{-}-Y^{+})(1-x^{+}Y^{-})}{2x^{+}(x^{+}-Y^{-})(1-x^{-}Y^{-})}\sqrt{\frac{Y^{-}}{Y^{+}}}\frac{x^{+}}{x^{-}}\displaybreak[0]\\
a_{9}^{9}(x^{\pm},Y^{\pm}) & =\frac{x^{-}-Y^{+}}{x^{+}-Y^{-}}\sqrt{\frac{Y^{-}}{Y^{+}}}\sqrt{\frac{x^{+}}{x^{-}}}\displaybreak[0]\\
a_{3}^{3}(x^{\pm},Y^{\pm}) & =\frac{x^{-}x^{+}(1+x^{-}Y^{-}-2x^{+}Y^{-})+\left(x^{+}+x^{-}(-2+x^{+}Y^{-})\right)}{x^{+}(x^{+}-Y^{-})(1-x^{-}Y^{-})}Y^{+}\sqrt{\frac{Y^{-}}{Y^{+}}}\sqrt{\frac{x^{+}}{x^{-}}}\\
 & \Rightarrow\quad\frac{a_{3}^{3}+a_{9}^{9}}{2}=\frac{(x^{+}-Y^{+})(x^{+}Y^{-}-1)}{(x^{+}-Y^{-})(x^{-}Y^{-}-1)}\sqrt{\frac{Y^{-}}{Y^{+}}}\sqrt{\frac{x^{+}}{x^{-}}}\,.
\end{align*}

At $M=1$ this reduces to the original $S_{1-1}$ (in our conventions) apart from
swopping bosons and fermions:
\[
\hat{S}_{M-1}(y^{\pm},x^{\pm})_{b1}^{b1}=\hat{S}(y^{\pm},x^{\pm})_{b'1}^{b'1},\qquad b'=b+2\mbox{ mod }4,\qquad M=1
\]
since clearly $a_{5}^{5}=a_{6}$, $a_{9}^{9}=a_{1}$ and we can show, using $M=1$,
that
\begin{align*}
\dfrac{a_{9}^{9}(x^{\pm},y^{\pm})+a_{3}^{3}(x^{\pm},y^{\pm})}{2} & =a_{1}(y^{\pm},x^{\pm})+a_{2}(y^{\pm},x^{\pm})\,.
\end{align*}

Then at $M=2$, given that the bound state is an A-particle plus a B-particle, the
full S-matrix will have prefactor $n\tilde{n}$, giving this: 
\begin{equation}
S_{b}(Y^{\pm},x^{\pm})=\hat{S}_{2-1}(Y^{\pm},x^{\pm})_{b1}^{b1}\; n(y_{1}^{\pm},x^{\pm})\tilde{n}(y_{2}^{\pm},x^{\pm})\;\sigma(Y^{\pm},x^{\pm})\,.\label{eq:S2-ABBN-with-prefactors}
\end{equation}

\subsection{Constructing $S_{2-Q}$ by fusion\label{sub:Constructing-S_2Q-by-fusion}}

For the dyonic second F-term we need to scatter a particular physical $Q$ bound
state (the dyonic magnon) with a virtual, mirror, $M=2$ bound state. We should be
able to construct the required $S_{2-Q}$ using $S_{2-1}$ from above. Let us write:
\begin{align*}
S_{2-Q}(Y^{\pm},X^{\pm})_{b1}^{b1} & =\prod_{k=1}^{Q}S_{2-1}(Y^{\pm},x_{k}^{\pm})_{b1}^{b1}\\
 & =T_{0}\: t_{b}\,.
\end{align*}

All the matrix elements are easy if we pull out a factor of $a_{9}^{9}$. Here is
the full list for $M=2$ (although clearly general $M$ would be no harder):
\begin{align*}
\qquad t_{1}=t_{2}=t_{3} & =\prod_{k=1}^{Q}\frac{a_{5}^{5}(x_{k}^{\pm},Y^{\pm})}{a_{9}^{9}(x_{k}^{\pm},Y^{\pm})}=\prod_{k=1}^{Q}\frac{Y^{+}-x_{k}^{+}}{Y^{+}-x_{k}^{-}}\sqrt{\frac{x_{k}^{-}}{x_{k}^{+}}} &  & =\frac{Y^{+}-X^{+}}{Y^{+}-X^{-}}\sqrt{\frac{X^{-}}{X^{+}}}\\
t_{4} & =\prod_{k=1}^{Q}\frac{2a_{8}^{8}}{a_{9}^{9}} &  & =\frac{\frac{1}{Y^{-}}-X^{+}}{\frac{1}{Y^{-}}-X^{-}}\sqrt{\frac{X^{-}}{X^{+}}}\\
t_{7}=t_{8} & =\prod_{k=1}^{Q}\frac{1}{2}\left(1+\frac{a_{3}^{3}}{a_{9}^{9}}\right) &  & =\frac{X^{-}}{X^{+}}\:\frac{\frac{1}{Y^{-}}-X^{+}}{\frac{1}{Y^{-}}-X^{-}}\:\frac{Y^{+}-X^{+}}{Y^{+}-X^{-}}\,.\qquad\qquad
\end{align*}
and $t_{5}=t_{6}=1$. (Note that for the cancellations obtained here, we do not need
to assume that the constituent particles have $Q=1$.)

For the scalar factor, insert $1=(y_{1}^{+}-x_{k}^{-})/(y_{2}^{-}-x_{k}^{-})$ so
as to create $s_{\mathrm{BDS}}=n/\tilde{n}$, giving the following cancellation:
\begin{align*}
t_{0,k} & =n(y_{1}^{\pm},x_{k}^{\pm})\:\tilde{n}(y_{2}^{\pm},x_{k}^{\pm})\: a_{1}(Y^{\pm},x_{k}^{\pm})\;\sigma(Y^{\pm},x_{k}^{\pm})\\
 & =s_{\mathrm{BDS}}(y_{1}^{\pm},x_{k}^{\pm})\:\frac{y_{2}^{-}-x_{k}^{+}}{y_{2}^{-}-x_{k}^{-}}\sqrt{\frac{x_{k}^{+}}{x_{k}^{-}}}\sqrt{\frac{Y^{-}}{Y^{+}}}\;\sigma(Y^{\pm},x_{k}^{\pm})\displaybreak[0]\\
\Rightarrow T_{0}=\prod_{k=1}^{Q}t_{0,k} & =S_{\mathrm{BDS}}(y_{1}^{\pm},X^{\pm})\:\frac{y_{2}^{-}-X^{+}}{y_{2}^{-}-X^{-}}\sqrt{\frac{X^{+}}{X^{-}}}\sqrt{\frac{Y^{-}}{Y^{+}}}^{Q}\;\sigma(Y^{\pm},X^{+})\,.
\end{align*}
This we can interpret as saying that, if $x^{\pm}$ is an A particle, then $y_{1}^{\pm}$
is A and $y_{2}^{\pm}$ is B. If instead it's the other way around, then we would
have written
\begin{align*}
\tilde{t}_{0,k} & =\tilde{n}(y_{1}^{\pm},x_{k}^{\pm})\: n(y_{2}^{\pm},x_{k}^{\pm})\: a_{1}(Y^{\pm},x_{k}^{\pm})\;\sigma(Y^{\pm},x_{k}^{\pm})\\
\Rightarrow\tilde{T}_{0}=\prod_{k=1}^{Q}\tilde{t}_{0,k} & =S_{\mathrm{BDS}}(y_{2}^{\pm},X^{\pm})\:\frac{y_{1}^{+}-X^{+}}{y_{1}^{+}-X^{-}}\sqrt{\frac{X^{+}}{X^{-}}}\sqrt{\frac{Y^{-}}{Y^{+}}}^{Q}\;\sigma(Y^{\pm},X^{+})\,.
\end{align*}
However it seems that one should not include both of these, as surely an A+B bound
state is the same as a B+A one. This gives the correct factors of 2 in the F-term
calculations (where we evaluate at $y_{k}^{\pm}=x+\mathcal{O}(1/h)$, and thus have
$T_{0}=\tilde{T}_{0}$). For $\mu$-terms, however, note that both $T_{0}$ and the
$t_{5}$ have poles at $Y^{-}=X^{+}$, while $\tilde{T}_{0}$ has instead a pole
at $Y^{+}=X^{+}$. 

Finally, we have also written down the elements $S_{2-Q}(Y^{\pm},X^{\pm})_{b2}^{b2}$
needed to treat the big giant magnon, in \eqref{eq:S_b2-M2-conjecture} above.

\section{Formulae for Magnons in the Algebraic Curve\label{sec:AC-Formulae}}

Since not all of the F-terms we calculated in sections \ref{sub:Dyonic-second-F-term}
and \ref{sec:F-terms-for-Composite-Magnons} were done explicitly in \cite{Abbott:2010yb},
we give here the formulae necessary to calculate these.

The algebraic curve for $AdS_{4}\times CP^{3}$ was introduced by \cite{Gromov:2008bz}.
The ansatz for magnons used by \cite{Lukowski:2008eq,Abbott:2009um}, but now including
twists needed to make a closed string as in \cite{Gromov:2008ec}, is as follows:
\begin{equation}
\begin{array}{ccccc}
q_{3}(x)= & \negmedspace\negmedspace\dfrac{\alpha x}{x^{2}-1} & \negmedspace\negmedspace\negmedspace\negmedspace\negmedspace\negmedspace+G_{u}(0)-G_{u}(\frac{1}{x})\tablespacer & +G_{v}(0)-G_{v}(\frac{1}{x})\quad & \negmedspace\negmedspace\negmedspace\negmedspace\negmedspace\negmedspace\negmedspace\negmedspace+G_{r}(x)-G_{r}(0)+G_{r}(\frac{1}{x})\;-\tau\\
q_{4}(x)= & \negmedspace\negmedspace\dfrac{\alpha x}{x^{2}-1} & \negmedspace\negmedspace\negmedspace\negmedspace\negmedspace\negmedspace+G_{u}(x)\tablespacer & +G_{v}(x)\quad & \negmedspace\negmedspace\negmedspace\negmedspace\negmedspace\negmedspace\negmedspace\negmedspace-G_{r}(x)+G_{r}(0)-G_{r}(\frac{1}{x})\;-\tau\\
q_{5}(x)= &  & \negmedspace\negmedspace\negmedspace\negmedspace\negmedspace\negmedspace\negmedspace\negmedspace G_{u}(x)-G_{u}(0)+G_{u}(\frac{1}{x}) & \negmedspace\negmedspace-G_{v}(x)+G_{v}(0)-G_{v}(\frac{1}{x})
\end{array}\label{eq:q12345-ansatz}
\end{equation}
and
\begin{align*}
q_{1}(x) & =q_{2}(x)=\dfrac{\alpha x}{x^{2}-1}\\
q_{n}(x) & =-q_{11-n}(x),\quad n=6,7,\ldots10\,.
\end{align*}
The residue at $x=\pm1$ is $\alpha=\Delta/h$. Giant magnons are made by turning
on 
\begin{equation}
G_{\mathrm{mag}}(x)=-i\log\left(\frac{x-X^{+}}{x-X^{-}}\right).\label{eq:defn-Gmag}
\end{equation}
in one or more slots:
\begin{itemize}
\item The elementary (or `small') giant magnon has 
\[
G_{v}(x)=G_{\mathrm{mag}}(x),\qquad G_{u}=G_{r}=0
\]
and $\tau=p/2$.
\item There is another kind of elementary magnon with $G_{u}$ instead of $G_{v}$. The
two correspond to A- and B-particles in the S-matrix language.
\item The $RP^{3}$ giant magnon is constructed by turning on both of these:
\[
G_{u}(x)=G_{v}(x)=G_{\mathrm{mag}}(x),\qquad G_{r}(x)=0.
\]
Here we should use $\tau=p$ since the total momentum is $2p$. 
\item The `big giant magnon' has 
\[
G_{u}(x)=G_{v}(x)=G_{r}(x)=G_{\mathrm{mag}}(x)
\]
and again $\tau=p$.
\end{itemize}
The factor in the energy correction which should match the supertrace of the S-matrix
we called $F$ in \cite{Abbott:2010yb}:
\[
F_{\mathrm{light}}^{(\ell)}=\sum_{\substack{ij\\
\mathrm{light}
}
}(-1)^{F_{ij}}e^{-i\ell(q_{i}-q_{j})},\qquad F_{\mathrm{heavy}}^{(\ell)}=\sum_{\substack{ij\\
\mathrm{heavy}
}
}(-1)^{F_{ij}}e^{-i\ell(q_{i}-q_{j})}.
\]
The list of heavy and light polarisations $(i,j)$ is as follows: \begin{equation}\label{eq-tab:heavy-and-light-ads-and-cp}%
\begin{tabular}{c|ccc}
 & $AdS$ & Fermions & $CP$\tabularnewline
\hline 
Heavy & (1,10) (2,9) (1,9) & (1,7) (1,8) (2,7) (2,8) & (3,7)\tabularnewline
Light &  & (1,5) (1,6) (2,5) (2,6) & (3,5) (3,6) (4,5) (4,6).\tabularnewline
\end{tabular}\end{equation}The fermions have $F_{ij}=1$ while the bosons have $F_{ij}=0$.

\begin{small}\bibliographystyle{my-JHEP-3}
\phantomsection\addcontentsline{toc}{section}{\refname}\bibliography{/Users/me/Documents/Papers/complete-library-processed,complete-library-processed}

\end{small}
\end{document}